\documentstyle[preprint,epsf,aps]{revtex}
\begin{document}
\tighten
\draft
    
\newcommand{\fig}[2]{\epsfxsize=#1\epsfbox{#2}}


\def\omn{\omega_{n}}
\def\epstar{\epsilon_F^{*}}
\def\ombar{\overline{\omega}}
\def\Tbar{\overline{T}}
\def\Si{\vec{S}_i}
\def\Sj{\vec{S}_j}
\def\gd{{\cal D}}
\def\Zeff{Z_{\hbox{\tiny single-site}}}
\def\intbeta{\int_{0}^{\beta}}
\def\intR{\int_{-\infty }^{\infty }}
\def\k{{\bf\mbox{k}}}
\def\q{{\bf\mbox{q}}}
\def\sgn{\mbox{sgn}}
\def\chiloc{\chi_{\mbox{\tiny loc}}}
\def\const{\mbox{\small const. }}
\def\LSCO{$\mbox{La}_{2-x}\mbox{Sr}_x\mbox{Cu}\mbox{O}_4$}
\def\YBCO{$\mbox{Y}\mbox{Ba}_2\mbox{Cu}_3\mbox{O}_{6+y}$}
\def\LBCO{$\mbox{La}_{1.95}\mbox{Ba}_{0.05}\mbox{Cu}\mbox{O}_4$}

\def\Im{\mathop{\hbox{Im}}}
\def\Re{\mathop{\hbox{Re}}}

\title{Non-Fermi liquid regime of a doped Mott insulator}

\author{Olivier Parcollet and Antoine Georges}
\address{Laboratoire de Physique Th{\'e}orique de l'Ecole Normale
Sup{\'e}rieure,
24 rue Lhomond, 75231 Paris Cedex 05, France}
\date{\today}
\maketitle

\begin{abstract}
We study the doping of a Mott insulator in the presence of
quenched frustrating disorder in the magnitude and sign
of the magnetic exchange.
Two quite different doping regimes $\delta<\delta^*$ and
$\delta>\delta^*$ are found, with $\delta^*\simeq J/t$
($J$ is the characteristic magnitude of the exchange,
and $t$ the hopping amplitude).
In the high-doping regime, a (Brinkman-Rice) Fermi liquid 
description applies with a coherence scale of order $\delta t$.  
In the low doping regime, local magnetic correlations 
strongly affect the formation of quasiparticles, resulting in 
a very low coherence
scale $\epsilon_F^* \simeq J (\delta/\delta^*)^2$.  
Fermi liquid behaviour does apply below $\epstar$, but a  
``quantum critical regime'' $\epsilon_F^*<T<J$ holds, in which 
{\it marginal Fermi liquid} behaviour of several physical
properties is found:  
NMR relaxation time $1/T_1\sim \mbox{const.}$, resistivity 
$\rho_{dc}(T) \propto T$, optical lifetime 
$\tau_{opt}^{-1}\propto \omega/\ln(\omega/\epstar)$ together with  
$\omega/T$ scaling of response functions, e.g. 
$J\sum_{\vec{q}}\chi''(\vec{q},\omega) \propto \tanh (\omega/2T)$.
 In contrast,  {\it single-electron} properties
display stronger deviations from Fermi liquid theory in this regime 
with a $\sqrt{\omega}$ dependence of the inverse single-particle 
lifetime and a $1/\sqrt{\omega }$ decay of the photoemission intensity. 

On the basis of this model and of various experimental evidence, 
it is argued that the proximity of a 
quantum critical point separating a glassy Mott-Anderson 
insulator from a metallic ground-state is an important ingredient in the 
physics of the normal state of cuprate superconductors.
In this picture the corresponding quantum critical regime is a ``slushy'' 
state of spins and holes with slow spin and charge dynamics  responsible for
the anomalous properties of the normal state. This picture may be particularly
relevant to  $Zn-$doped materials.
\end{abstract}

\pacs{PACS: 71.10.Hf, 71.30.+h, 74.72.-h. Preprint  LPTENS 98/24}

\widetext

\section{Introduction}
\label{intro}

How (and whether) coherent quasiparticles form in a lightly doped 
Mott insulator is a key question in the physics of strongly 
correlated electron systems. A satisfactory theoretical understanding 
of this issue has been achieved in the limit where magnetic correlations 
do not play a prominent role, starting with the work of 
Brinkman and Rice \cite{BR,largen,review}. 
In cuprate superconductors however, the 
undoped phase is an antiferromagnetic insulator with a rather large 
exchange coupling $J_{AF}$ (on the scale of $100\,meV$), 
so that we have to face the problem of the 
interplay between local coherence and magnetic correlations. 

Furthermore, 
there is ample experimental evidence that {\it carrier localisation} and
{\it magnetic frustration} also 
play a crucial role in the low to intermediate doping regime. 
This is particularly clear in the \LSCO\ compound at 
concentrations just above $x=0.02$ (the threshold for the disappearance of the 
antiferromagnetic long-range order), for   
which true spin-glass ordering of the copper moments has been 
demonstrated at very low temperature (with $T_g\simeq 7K$ for 
$x=0.04$ \cite{sglsco}). 
Up to which doping concentration does this glassy regime 
persist when superconductivity is suppressed is not known at this 
point, but carrier localization is indeed observed at low temperature 
up to optimal doping  
in both the $ab$ and $c$ directions when a strong magnetic field is 
applied \cite{ando1,boeb}. 
It was actually predicted early on \cite{aha} that hole doping induces 
strong frustration in the system when the holes become localized,
replacing locally an antiferromagnetic 
Cu-Cu bond with an effectively {\it ferromagnetic} one, with a strength 
larger than the original $J_{AF}$. We observe furthermore that the 
disappearance of antiferromagnetic long-range order is accompanied by 
the appearance of new low-energy spin excitations, of a quite different 
nature than spin waves, as evidenced by inelastic neutron 
scattering experiments \cite{nlabaco}-\cite{sidis}. 
It is  important to notice that the 
compounds with a glassy ground-state display, at sufficiently 
high temperature (above the onset of localization), 
the same distinctive transport properties as in 
samples with higher doping, e.g. linear resistivity \cite{nlabaco,nlscosg}. 
It is thus tempting to view these 
low-energy excitations as the source of anomalous scattering in 
the normal state.  

Anticipating  some of the speculations made at the end of this paper, 
we shall argue that these low energy excitations are associated with 
a novel kind of spin state: the ``slushy'' state associated with the 
disordering of an insulating (possibly glassy) ground state by hole 
motion, quantum fluctuations and thermal effects. In this picture, 
many distinctive 
``anomalous'' properties of the normal state of the cuprate 
superconductors are associated with the quantum 
critical regime corresponding to the $T=0$ transition at which 
the insulating (glassy) ground-state melts into a metallic (Fermi-liquid)
ground state when doping is increased.

In this paper, we shall study a highly simplified model of such 
a slushy state of spins and holes. Our starting point is the 
work of Sachdev and Ye \cite{spinfluid}, who showed  
that in the large-$M$ limit of the fully-connected random 
Heisenberg model of $SU(M)$ spins, quantum fluctuations are strong 
enough to overcome the tendency to spin-glass ordering. 
Instead, a gapless spin-liquid state is found down to zero
temperature with a large density of low-energy spin excitations
\cite{warningbosons}. 
Remarkably, these excitations are characterized by a local dynamic spin 
susceptibility which has precisely the form advocated by the ``marginal 
Fermi liquid'' phenomenological description \cite{mfl} 
of the low energy spin excitations in cuprates, namely:
\begin{equation}\label{mflt=0}
\chi_{loc}''(\omega,T=0) = {{\pi}\over{2J}}\mbox{sgn}(\omega)\,\,\,,\,\,\,
\chi_{loc}'(\omega,T=0)={{\sqrt{\pi}}\over{2J}} 
\ln \left({{J}\over{|\omega|}}\right)
\end{equation}
This model is one of the few cases in which a response 
function having the marginal Fermi liquid form could be derived
explicitly (see also \cite{nestedfl}).
 The generalization of Eq.(\ref{mflt=0}) to
finite temperature will be given in Sec.\ref{SectCrossOverTherm} 
(Eq. (\ref{MflChi}))
and displays $\omega/T$ scaling. 
The physical mechanism for the gaplessness and the high-density of
 spin excitations in this model  
is discussed in more detail at the beginning of Sec.\ref{physics}. 
It has to do with the large number of transverse components of the 
spins in the large-$M$ limit. In this respect, it might be a
reasonable picture for the disordering of the two-dimensional
quantum Heisenberg spin-glass due to quantum fluctuations and low 
dimensionality \cite{grempel}. 

The main purpose of this paper is to determine whether this marginal 
Fermi liquid spectrum survives the introduction of charge carriers 
and the associated insulator to metal transition. 
The physics of this problem is dominated by the interplay between two 
competing effects:
\begin{itemize}

\item The formation of coherent metallic quasiparticles, which can be 
viewed as a binding of spin and charge degrees of freedom. In the simplest 
description of a doped Mott insulator with $U=\infty$, 
coherent quasiparticles form below a 
scale of order $T^*_{F0}\sim\epsilon^*_{F0}\sim \delta t$ (where $\delta$ 
is the doping and $t$ the hopping amplitude). This is a ``naive'' estimate of 
the effective Fermi-energy scale, since it ignores any effect
coming from the magnetic exchange (which will tend to suppress it).

\item The binding of spin degrees of freedom on neighbouring sites into 
singlet or triplet states, and the corresponding slow dynamics of the 
on-site local moment. 
This is the phenomenon leading to the formation of the spin-liquid state 
in the undoped phase, which involves a scale of order $J$ (the characteristic 
strength of the exchange). 

\end{itemize}

\begin{figure}
\[
\fig{10cm}{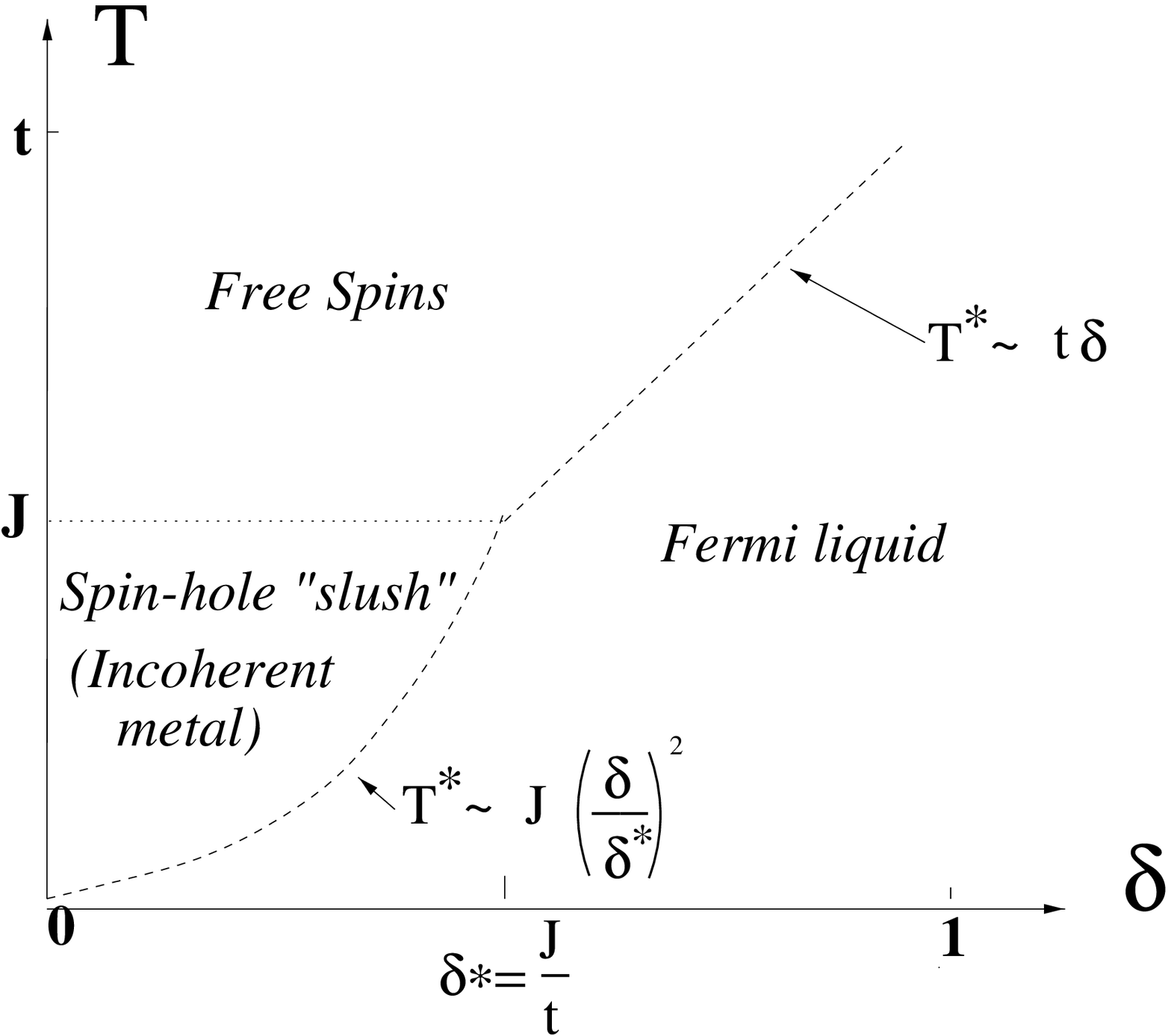} 
\]
\vskip 1cm
\caption{\label{Crossovers}
Crossover diagram as a function of temperature and doping. The coherence scale
$\epstar $ is indicated by a dashed line and is given by 
$\epstar \simeq  J (\delta /\delta^{*})^{2}$ for $\delta <\delta^{*} $, 
$\epstar \simeq \delta  t $ for $\delta > \delta^{*}$, with $\delta^{*} =
J/t$. Below $\epstar $, Fermi liquid behaviour holds. For $\delta
<\delta^{*}$, an intermediate ``quantum critical'' regime is found in the
range $\epstar <T <J$, in which charge transport is incoherent
  and spins have a marginal Fermi liquid dynamics.
}
\end{figure}

It is clear from comparing the scales above that when $J$ is larger than 
the ``naive'' coherence scale $\epsilon^*_{F0}$, the magnetic exchange 
prevents the formation of coherent quasiparticles at that scale: in other 
words, $\epsilon^*_{F0}$ cannot possibly be the {\it actual quasiparticle 
coherence scale}  
above which free local moments are recovered, since the exchange is 
still effective at energy scales between $\epsilon^*_{F0}$ and $J$. It is 
thus expected that the {\it actual} coherence scale of the system, 
$\epsilon^*_{F}$ will 
be much smaller than $\epsilon^*_{F0}$, and that a new metallic regime 
in which 
spin degrees of freedom form  a spin-liquid like state while 
charge degrees of freedom are incoherent will be found in the intermediate 
energy and temperature range $\epsilon^*_{F}< \omega,T < J$. 
From the above estimates, this will be the case at small doping: 
$\delta<\delta^*\sim J/t$, while a direct crossover from a coherent 
metal to 
an incoherent high temperature state is expected for $\delta>\delta^*$. 
These expectations are entirely borne out from our solution of the doped 
Sachdev-Ye model, as evidenced by Fig.\ref{Crossovers}, which 
summarizes the main crossovers found in our analysis. 

It should be emphasized that this competition between metallic 
coherence and magnetic exchange is also essential to 
the physics of heavy fermion compounds \cite{doniach}. 
In this context, the ``naive'' coherence scale $\epsilon^*_{F0}$ stands for 
the single-impurity Kondo scale (or rather, any estimate of the lattice 
Kondo scale that ignores RKKY interactions), while $J$ stands for the 
typical strength of the RKKY interaction. 
For this reason, the results of the present paper may also have some 
relevance, with appropriate changes, to the physics of the disordered 
rare-earth compounds near the quantum critical transition into a spin glass
ground state \cite{sghf}.

\section{The Model}
\subsection{Disordered $SU(M)$ $t-J$ model}

The effect of charge carriers on the Sachdev-Ye spin-liquid phase will be
investigated by generalizing the model of Ref. \cite{spinfluid} to 
a $t-J$ model, 
with randomness on the exchange couplings $J_{ij}$ between nearest neighbors
sites : 
\begin{equation}\label{Ham1}
H = - \sum_{<ij>\alpha}t_{ij}\, P \, c^{\dagger}_{i\alpha}c_{j\alpha} \, P + 
\sum_{<ij>} J_{ij} \vec{S}_{i}\cdot\vec{S}_{j}
\end{equation}
In this expression, the $SU(2)$ spin symmetry of the electrons has been
enlarged to  $SU(M)$ \cite{SUN}.
$\vec{S}_i$ is the conduction electron spin 
density on site $i$  and the spin index $\alpha $ runs over 
$\alpha=1,\cdots,M$.
The projection operator $P$ enforces the local constraint:
\begin{equation}\label{Contrainte1}
\sum_{\alpha}c^{\dagger }_{i\alpha}c_{i\alpha} \leq \frac{M}{2}
\end{equation}
In this manner the $M=2$ case exactly coincides with the standard $t-J$ model
with the  constraint of no double occupancy.

The exchange couplings are quenched random variables with random 
sign and magnitude, distributed according to a Gaussian distribution with :
\begin{equation}\label{DefJij}
J_{ij}= {J \over\sqrt{zM} } \epsilon_{ij}     \,\,\,,\,\,\,
\overline{\epsilon_{ij}}=0                           \,\,\,,\,\,\,
\overline{\epsilon_{ij}^2} = 1
\end{equation}
(throughout this paper the bar will denote an average over the disorder).
In the following, we shall consider this model on a lattice of 
connectivity $z$, with a nearest-neighbor hopping amplitude  
normalized as:
\begin{equation}\label{Deftij} 
t_{ij} = {{2t}\over{M\sqrt{z}}}
\end{equation}
and we shall analyze the model in the following double limit: 
\begin{itemize}
\item [i)]  $z\rightarrow\infty$. In this limit of infinite connectivity, 
a dynamical mean field theory applies which reduces the model to 
the study of a single-site self-consistent problem \cite{review}, as detailed 
in Sec. \ref{ReductionUnSite}. However this single site model is still
a complicated interacting problem. For the sake of simplicity, the lattice 
will be taken to be a Bethe lattice (no essential physics is lost in this 
assumption).          
\item [ii)]  $M\rightarrow \infty $, in which the single site 
problem becomes tractable. In the absence of a random exchange,
this limit yields the familiar
Brinkman-Rice description of a doped Mott insulator \cite{BR,SUN}.
\end{itemize}
The scaling in $z$ and $M$ in (\ref{DefJij}) and (\ref{Deftij}) 
are chosen such that this double limit gives non trivial results.
Alternatively, one could consider (as in \cite{spinfluid})
 this model on a fully connected lattice
 of $N$ sites, with {\it random} hopping amplitudes: 
$t_{ij}={{2t}\over{M\sqrt{N}}}\xi_{ij}$ 
with $\overline{\xi_{ij}}=0$, $\overline{\xi_{ij}^2}=1$.
This leads to precisely the same equations for single particle Green's 
functions as the $z=\infty$ Bethe lattice \cite{review}. 

We shall use  a decomposition of the physical electron
 operator into a spin-carrying fermion $f$ and a slave boson $b$ :
 $c^+_{i\alpha}=f^+_{i\alpha}b_i$. The local constraint (\ref{Contrainte1})
becomes~:
\begin{equation}\label{DefConstraint}
\sum_{\alpha} f^{\dagger}_{i\alpha}f_{i\alpha} + b^{\dagger}_{i}b_{i}
 = {M \over 2}
\end{equation}
With this decomposition the Hamiltonian (\ref{Ham1}) 
can be rewritten as:
\begin{equation}\label{Ham2}
H =  - {{2t}\over{M\sqrt{z}}}
\sum_{<ij>\alpha} (f^{\dagger}_{i\alpha}b_i b^{\dagger}_j f_{j\alpha} + h.c)  
+ {{J}\over{\sqrt{Mz}}}
\sum_{<ij>}\epsilon_{ij} 
\sum_{\alpha\beta} S_{i\alpha\beta}S_{j\beta\alpha}
\end{equation}
and the $M^2-1$ components of the $SU(M)$ spin operators 
$\vec{S}_i=(S_i)^{\alpha\beta}$ read : 
\begin{equation}\label{ExprSf}
S_{i\alpha\beta}=f^+_{i\alpha}f_{i\beta}-{{1}\over{M}}
\delta_{\alpha\beta} 
\sum_{\alpha}f^+_{i\alpha}f_{i\alpha}
\end{equation} 

\subsection{Reduction to a single-site problem}\label{ReductionUnSite}

In this section, we explain how the large connectivity limit 
$z\rightarrow \infty$ reduces the problem to the study of a single-site 
model supplemented by a self-consistency condition.  
First we use a path integral representation  
of the partition function $Z$ and introduce a Lagrange multiplier 
field $\lambda_i(\tau)$ on each site in order  
to handle the constraint (\ref{DefConstraint}). 
We then introduce $n$ replicas of the fields 
($f_{i}^{a},b_i^{a},\lambda_i^a\,\,\,,a=1,\cdots,n$) in order to express 
$Z^n$ and average over the disorder. 
The action associated with $\overline{Z^n}$ reads: 

\begin{eqnarray}\label{action1}
\nonumber
S = &&\sum_{i,a}  S_{0}[f^{\dagger a}_{i},f_{i}^{a},
b^{\dagger a}_{i},b_{i}^{a},\lambda_{i}^{a}] 
 - {2t\over\sqrt z M} \sum_{<i,j>,\alpha}\int_0^{\beta} d\tau  
f^{\dagger a}_{i\alpha}(\tau) 
b^a_i(\tau) b^{\dagger a}_j(\tau) f^a_{j\alpha}(\tau) \\
&&-{ J^2\over 2 zM}\int_0^{\beta}\!\!\!\int_0^{\beta} d\tau d\tau'\, 
\sum_{<ij>}\sum_{1\leq\alpha,\beta,\gamma,\delta\leq M} 
\sum_{1\leq a,b\leq n} \,  
 S^{a}_{i\alpha \beta } (\tau)   S^{a}_{j\beta \alpha }  (\tau)
 S^{b}_{i\gamma \delta }(\tau')  S^{b}_{j\delta \gamma } (\tau')  
\end{eqnarray}

where the action $S_{0}$ is defined by
\begin{eqnarray}\label{action0}
\nonumber 
S_{0}[f^{\dagger},f,b^{\dagger},b,\lambda]
&\equiv &  \int_0^{\beta} d\tau \left( 
b^{\dagger }(\tau)\partial_\tau b(\tau) + 
\sum_{\alpha} f^{\dagger}_{\alpha}(\tau)(\partial_\tau-\mu )
 f_{\alpha}(\tau) \right) \\
&& + i \intbeta d\tau \lambda(\tau)
\left(
\sum_{\alpha} f^{\dagger }_{\alpha}(\tau) f_{\alpha}(\tau) + 
b^{\dagger }(\tau)    b(\tau)         - \frac{M}{2}
\right) 
\end{eqnarray}
Following the ``cavity method'' (reviewed in \cite{review}), a 
site of the lattice is singled out, and a trace is performed over 
all degrees of freedom at the other sites (concentrating on phases 
without translational symmetry breaking, so that all sites are 
equivalent). In the $z\rightarrow\infty$ limit, this can be performed 
explicitly, and the problem reduces to a single-site effective action which 
reads:  
\begin{eqnarray}\label{action2}
\nonumber
S_{eff}=&&\sum_{a} 
S_{0}[f^{\dagger a},f^{a},b^{\dagger a},b^{a},\lambda^{a}] 
- { J^2 \over 2 M} \sum_{a,b,\alpha ,\beta ,\gamma ,\delta }
\intbeta\!\!\!\intbeta  d\tau d \tau'\,
S^{a}_{\alpha \beta }(\tau) 
R^{ab}_{\beta \alpha  \delta \gamma  }(\tau-\tau')
S^{b}_{\gamma \delta }(\tau') \\
&&+\left( \frac{2t}{M} \right)^{2}
\sum_{a,\alpha}\intbeta\!\!\!\intbeta d\tau d \tau'\,  
 f^{\dagger a}_{\alpha}(\tau) b^a(\tau)
 C^{aa}_{\alpha\alpha}(\tau-\tau')
 b^{\dagger a}(\tau')  f^a_{\alpha}(\tau')
\end{eqnarray}
This effective action is supplemented by a self-consistency condition 
which constrains $C(\tau-\tau')$ and $R(\tau-\tau')$  
to coincide with the local electron Green's function and spin correlation 
function respectively:  
\begin{eqnarray}\label{defDetR}
\nonumber
C^{aa}_{\alpha\alpha}(\tau,\tau')&=&
-
{\left\langle T
c^{a}_{i \alpha}(\tau) c^{\dagger a}_{i\alpha} (\tau')
\right\rangle}_{S}
= -
{\left\langle T
(f^{a}_\alpha b^{\dagger a})(\tau)
(f^{\dagger a}_\alpha b^{a})(\tau')
\right\rangle}_{S_{eff}} \\
 R^{ab}_{\alpha \beta \gamma \delta }(\tau,\tau') &=& 
{\left\langle 
S^{a }_{i\alpha \beta }(\tau) S^{b}_{i\gamma \delta }(\tau')
\right\rangle}_{S}
= 
{\left\langle 
S^{a}_{\alpha \beta}(\tau) S^{b}_{\gamma\delta}(\tau')
\right\rangle}_{S_{eff}}
\end{eqnarray}
In each of these equations, the last equality expresses the fact that 
local correlation functions can be calculated using the single-site action 
$S_{eff}$ itself. The limit $n\rightarrow 0$ must eventually be taken 
in these equations.
 
\subsection{Saddle-point equations in the large-$M$ limit 
and slave-boson condensation}\label{LargeMLimit}

We shall study the above self-consistent single-site problem in the large 
$M$ limit, focusing on the paramagnetic phase of the model. 
In this case, all the above
correlators become 
replica diagonal ($C^{aa}=C$, $D^{ab}=D\delta_{ab}$). 

Furthermore, we shall look for solutions in which the slave boson undergoes 
a Bose condensation. (Solutions with an uncondensed boson when the bosons
carry an additional channel index have been investigated by Horbach and 
Ruckenstein \cite{HR}).
The solutions considered here  can be found as a saddle-point of 
$S_{eff}$ by setting: $b(\tau)=\sqrt{M/2} \phi(\tau)$ and looking for 
solutions in which 
both $\phi(\tau)$ and the Lagrange multiplier $\lambda(\tau)$ become 
static at the saddle-point:
\begin{equation}
b_{sp}(\tau) = \sqrt{M\over 2} \sqrt{\delta}\,\,\,,\,\,\,
i\lambda_{sp}(\tau) = \lambda_0
\end{equation}
From the constraint Eq.(\ref{DefConstraint}), the total number of electrons 
will be related to $\delta$ through: 
$\sum_{\alpha}<f^\dagger_{\alpha}f_{\alpha}>={M\over 2} (1-\delta)$ 
so that $\delta$ measures the number of holes doped into the system. 
   
The saddle point equations then reduce to a non-linear integral equation for 
the fermion Green's function 
$-<Tf_{\alpha}(\tau)f^\dagger_{\beta}(\tau')>\equiv 
\delta_{\alpha\beta}G_f(\tau-\tau')$, which reads 
(with : $\omn = (2n+1)\pi /\beta $ the Matsubara frequencies):
\begin{mathletters}\label{EqBase}
\begin{equation}\label{EqBase1}
G^{-1}_{f}(i\omn)= i\omn+\mu -\lambda_0 -(t\delta)^2 G_{f}(i\omn) 
-\Sigma_{f}(i\omn)
\end{equation}
\begin{equation}\label{EqBase2}
\Sigma_{f}(\tau) \equiv  -J^2 G^2_{f}(\tau) G_{f}(-\tau)
\end{equation}
and to the following relations, which determine the Lagrange multiplier 
$\lambda_0$ and the chemical potential $\mu$ for given values of the 
doping $\delta$ and the temperature (given the couplings $J$ and $t$): 
\begin{equation}\label{EqBase3}
G_{f}(\tau=0^-) = \frac{1-\delta }{2}
\end{equation}
\begin{equation}\label{EqBase4}
\lambda_0 \sqrt \delta = -2 t^2 \delta^{\frac{3}{2}}
 \int_{0}^{\beta} \! d \tau  G_{f}(\tau)G_{f}(-\tau)
\end{equation}
\end{mathletters}
The derivation of these saddle point equations from $S_{eff}$ is 
detailed in Appendix \ref{apxspeq}. 

The {\it local} spin-spin correlation function is directly 
related to $G_f$ in the $M\rightarrow\infty$ limit, as:
\begin{equation}
R(\tau) \equiv {{1}\over{M^2}}\sum_{\alpha\beta} 
<S_{i\alpha\beta}(\tau)S_{i\beta\alpha}(0)> = - G_f(\tau)G_f(-\tau)
\end{equation} 
In the following, we shall often consider the spectral functions  
associated with the single-particle Green's function and the 
local spin-spin correlation:
\begin{equation}\label{DefRho}
\rho_{f}(\omega) \equiv - \frac{1}{\pi} \Im G_{f}(\omega +i0^{+})
\,\,\,,\,\,\,
\chi_{loc}''(\omega) \equiv  \Im R(\omega +i0^{+})
\end{equation}

\section{Physical properties of the metallic state}
\label{physics}

In this section, we study the nature of the metallic state as a 
function of the doping level $\delta$.  

\begin{figure}
\[
\fig{10cm}{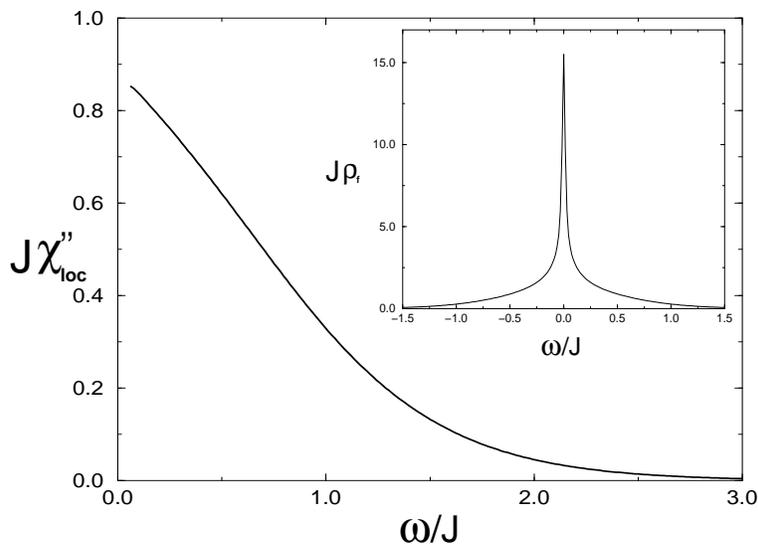} 
\]
\vskip 1cm
\caption{\label{DosSpinFluid}
Local dynamical susceptibility $\chiloc''(\omega,T=0 )$ of the undoped spin
liquid. Inset :  spectral function.}
\end{figure}

Let us first recall some of the properties of the spin-liquid insulating 
state found for $\delta=0$, as obtained by Sachdev and Ye 
\cite{spinfluid}. In this case, our equations 
(\ref{EqBase1}-\ref{EqBase3}) coincide with those of Ref.\cite{spinfluid}.  
Note that Eq.(\ref{EqBase4}) decouples, being automatically satisfied 
at $\delta=0$, and that particle-hole symmetry imposes 
$\mu-\lambda_0=\Sigma_f'(i0^+)=0$. 
A low-frequency analysis of the integral equation reveals that the $T=0$
Green's 
function and spectral density have a $1/\sqrt{\omega}$ singularity for 
$|\omega|\rightarrow 0$. More precisely \cite{footsubir}, 
in the complex frequency plane as $z\rightarrow 0$:
\begin{equation}\label{LowOmBehavG}
G_f(z) = \left({{\pi}\over{4J^2}}\right)^{1/4} 
{{(1-i)}\over{\sqrt{z}}} + \cdots \,\,\,,\,\,\,
\mbox{Im}z > 0
\end{equation}
This yields the following behaviour of the local dynamical susceptibility 
for $\omega\rightarrow 0$:
\begin{equation}
\label{mflchi}
\chi_{loc}''(\omega) = {{\pi^{3/2}}\over{4J}}\, \mbox{sgn}(\omega) + \cdots
\end{equation}
Fig.\ref{DosSpinFluid} displays a numerical calculation of 
$\rho_f(\omega)$ and $\chi_{loc}''(\omega)$  
at zero doping (in agreement with the one in Ref.\cite{spinfluid}). 
These results  display the above low-frequency behaviour (but we 
note that significant corrections to (\ref{mflchi}) are already sizeable at
rather low values of $\omega /J$.)

Hence the insulator at $\delta=0$ is a gapless quantum paramagnet 
(spin liquid), with a 
rather large density of low-energy spin excitations. 
Remarkably, (\ref{mflchi}) is of the same form than the ``marginal Fermi 
liquid'' susceptibility proposed on phenomenological grounds by Varma 
{\it et al.} \cite{mfl} for the normal state of the cuprate 
superconductors. In the present context, the physical nature of these 
low-energy excitations is intimately connected to the fact that the 
exchange couplings $J_{ij}$ are random in sign. In constructing the 
ground-state of the insulator, let us imagine that we first try to 
satisfy the bonds with the larger exchange constants. When such a 
bond is antiferromagnetic, the two spins connected by it will form 
a non-degenerate singlet. For a ferromagnetic bond however, the 
two spins will pair into a state of maximal possible spin (the 
generalization to $SU(M)$ of a triplet state). This state has a degeneracy, 
which actually becomes very large (exponential in $M$) as $M$ becomes 
large. Continuing the process in order to accommodate bonds with 
smaller strengths will tend to remove part of this degeneracy 
\cite{footentropy}, but leaves 
behind a very large density of low-energy spin excitations. 
These effects are clearly favored by the (fermionic) large-$M$ limit 
considered here, because of the high degeneracies of the ``triplet'' state 
and because the strength of quantum fluctuations in this limit precludes 
the appearance of long-range order (e.g. spin-glass)
 which would remove degeneracies in a 
different manner. 
We believe however that  this physics 
is not  an {\it artefact} of the large$-M$ limit.
Indeed,  preliminary theoretical studies \cite{Anirvan}
 suggest that the local spin 
correlations near the quantum critical point associated with the 
$T=0$ transition into a metallic spin-glass phase could be similar 
to Eq.(\ref{mflchi}), with a related physics.

Finally, we note that the single-site action to which the model reduces 
at zero doping
({\it i.e} Eq. (\ref{action2}) with $t=0$) has some similarities 
with the multichannel Kondo effect in the {\it overscreened} case. 
In the present context however, the ``bath'' seen by the spin is not 
due to an electronic conduction band, but generated by all the other 
spins in the lattice. The spin correlations of both the bath and the 
spin adjust to the self-consistent long-time behaviour: 
$<S(0)S(\tau)> \sim R(\tau) \sim 1/\tau$ similar to that of the 
$SU(M)$ Kondo model with $K=M$ channels \cite{OPAGKondo} 
(2-channel model in the $SU(2)$ case).   
    
\subsection{Low-frequency analysis and the Fermi liquid coherence scale}
\label{LowFreqAnalysis}
The first question we would like to address is whether the ``marginal 
Fermi liquid'' spin dynamics survives the introduction of charge carriers. 
As we shall demonstrate, this depends on the temperature range 
considered (Fig.\ref{Crossovers}). 
At low temperature, below some -possibly very low- 
coherence scale $\epsilon_F^*$, it turns out that a Fermi liquid is 
recovered. 

This is easily seen from a low-frequency analysis of the integral equation 
for $G_f$ at zero-temperature. At zero doping, the Green's function and 
self-energy behave at low-frequency as:
$G_f(\omega)\sim 1/\sqrt{J\omega}\,\,,\,\,
\Sigma_f(\omega)\sim \sqrt{J\omega}$. When inserted in 
Eq.(\ref{EqBase1}), this controls the leading low-frequency behaviour 
of both the r.h.s and l.h.s of the equation taken at $\delta=0$, 
which match each other. However, for $\delta\neq 0$, the term 
$(t\delta)^2 G_f(\omega)$ would introduce a $1/\sqrt{\omega}$ singularity 
and prevent this matching from taking place: this 
indicates that the low-frequency behaviour of the zero-temperature 
Green's function for arbitrary small doping is no longer 
$1/\sqrt{\omega}$. In this respect, 
an infinitesimal doping is a {\it singular perturbation} of 
the above equations. This observation directly yields an estimate of the 
coherence scale $\epsilon_F^*$ such that $G_f(\omega)\sim 1/\sqrt{\omega}$ 
is recovered for $\epsilon_F^*<\omega\ll J$. Indeed, the term 
$(t\delta)^2 G_f(\omega)$ becomes comparable to $\Sigma_f(\omega)$ in this 
regime (thus providing a cut-off to the singular behaviour) when 
$\omega\simeq (\delta t)^2/J$. Hence, in the low-doping regime:
\begin{equation}
\label{scalesmall}
\epsilon_F^* = {{(\delta t)^2}\over{J}}\,\,\,,\,\,\,
(\delta\ll\delta^*)
\end{equation}
where $\delta^*$ will be precised below. (In the following we shall 
take Eq.(\ref{scalesmall}) as defining  $\epsilon_F^*$ in 
the low-doping regime, with no additional prefactors).
 
In the high-doping regime on the other hand (or when $t\gg J$), one should 
consider first the limit of a vanishing magnetic exchange $J=0$. 
In this limit, the usual slave-boson (large $M$) description of a doped 
Mott insulator is recovered \cite{largen}.
Setting $J=\Sigma_f=0$ in the 
equations above yields a semi-circular spectral density:
\begin{equation}
\rho_f^{J=0} = \frac{1}{\delta } 
D\left(\frac{\omega+\mu-\lambda_0}{\delta}\right)
\label{dosj=0}
\end{equation}
where $D$ is given by :  
\begin{equation}\label{DefD}
D(\epsilon ) = \frac{1}{\pi t}\sqrt{1-\left(\frac{\epsilon}{2t} \right)^{2}}
\end{equation}
The original bandwith $4t$ of the non-interacting case has been 
reduced by a factor $\delta$, and the usual Brinkman-Rice result for the 
coherence scale is recovered :
\begin{equation}
\label{scalelarge}
\epsilon_F^* = t\delta  \,\,\,,\,\,\, (\delta\gg\delta^*)
\end{equation}
Turning on $J$ as a perturbation from this starting point does not affect 
the leading low-frequency behaviour of the self-energy, but does lead to
a scattering rate
$\Im\Sigma_f\propto \omega^2 J^2/(\delta t)^3 +\cdots$ characteristic of
a Fermi liquid (in contrast the $J=0$ model has infinite quasiparticle
lifetime  in the large-$M$ limit). From Eqs.(\ref{scalelarge},\ref{scalesmall}),
it is clear that when the  magnetic scattering is strong ($J\gg t$), regime
(\ref{scalesmall})
always applies, while for weaker scattering ($J<t$) a crossover
between the two regimes is found at a characteristic doping:
\begin{equation}
\delta^* \sim \mbox{min}({{J}\over t}, 1)
\end{equation}
We thus observe that below some characteristic doping the low-energy
coherence scale is strongly affected by the magnetic scattering.
When the exchange is large  or for doping smaller that
$\delta^*\simeq J/t$, 
the actual coherence scale $\epstar $ is  much smaller than
the ``naive'' coherence scale $\epsilon^*_{F0}$ (which holds in the absence of
magnetic correlations). 
Here we find $\epsilon^*_{F0} \simeq \delta t $ and 
$\epsilon_F^*/\epsilon^*_{F0}\simeq \delta/\delta^*$. 
This is one of the crucial physical conclusions of this paper.

A numerical solution of the saddle-point equations provide a clear evidence
for these two regimes. The numerical procedure that we have used is 
explained in Appendix \ref{AppendiceNum}. 
Fig. (\ref{SpectralFunctions}) displays the
$T=0$ spectral function $\rho_{f}(\omega)$
for three values of $J$. When $J$ is very small, 
the spectral function is very close to the
semi-circular shape (\ref{dosj=0}), 
while for a larger $J$ the $1\over \sqrt{\omega}$ divergence 
is observed over a large frequency range $\epsilon_F^*<\omega<J$ but is
cutoff for $\omega<\epsilon_F^*$ so that $\rho_f(0)$ is finite. Anticipating
on the results of Sec. \ref{PropUnePart}, we observe that the value of 
$\rho_f(\omega=0,T=0)$ is
actually independent of $J$ as a consequence of the 
Luttinger theorem. Indeed, the
following relation can be established at zero-temperature:
\begin{equation}
\label{ThmLuttinger}
\mu-\lambda_0-\Sigma_f(i0^+)\rightarrow \delta\mu_0(\delta)
 \qquad \qquad  \mbox{as } \qquad  
T\rightarrow 0
\end{equation}
where $\mu_0(\delta)$ is the non-interacting value of the chemical 
potential for the tight binding model on the $z=\infty$ Bethe lattice. 
This implies:
$\rho_f(0, T=0)=1/(\pi t\delta)$ for all values of $J$.

At very small doping, a scaling analysis of the saddle-point equations 
can be performed in order to characterize more precisely the crossover
between the low-frequency and high frequency regimes at $T=0$. 
As we now show, the spectral function (and the Green's function itself)
 obeys a scaling form:
\begin{equation}\label{DefPhif}
\rho_{f}(\omega) = \frac{1}{t\delta }\,
 \phi_{f}\left(\frac{\omega }{\epstar } \right)
\,\,\,\,\, \hbox{for} \,\,\,\,\,
\omega  \ll J,t   \,\,\,\,\,\,\,\,\,  \delta \ll \delta^*=\frac{J}{t}
\end{equation}
In order to derive the integral equation satisfied by the scaling function
$\phi_{f}$, we rewrite Eq.(\ref{EqBase1}) at $T=0$ (using (\ref{ThmLuttinger}))
as :
\begin{equation}\label{lowdoping1}
G_{f}^{-1}(\omega ) = \omega  + \delta \mu_{0}(\delta ) - 
(t\delta)^{2} G_{f}(\omega ) -
 \left(\Sigma_{f}(\omega ) - \Sigma_{f}(0) \right)
\end{equation}
At low doping and low-frequency $G_f$ is of order $1/\delta$ and
$\mu_0$ is of order $\delta$. Hence, rescaling frequencies by the
coherence scale $\epsilon_F^* = (\delta t)^2/J$, we see that
the first two terms in the r.h.s of (\ref{lowdoping1}) 
can be neglected.
Analytically continuing to {\it real} time $t$ and 
frequency $\omega$, and denoting  by $G^{F}$ the real-frequency
 $T=0$ Green's function (with the usual Feynman prescription), 
we define a scaling function  $g_{f}^{F}$ associated with  $G_f^{F}$ by  
the Hilbert transform:
\begin{equation}
g_{f}^{F}(\omega) = \int_{-\infty }^{\infty } d\epsilon \,
\frac{\phi_{f}(\epsilon)}{\omega -\epsilon + i \,\sgn \omega }
\end{equation}
We finally  obtain from (\ref{lowdoping1}) 
an integral equation satisfied by $g_f$ (and thus by $\phi_{f}$) which
 no longer contains dimensional parameters: 
\begin{eqnarray}\label{Equationgf}
\nonumber
(g_{f}^{F}(\omega))^{-1} 
&=& - g^{F}_{f}(\omega) -\left(\sigma^{F}(\omega)-\sigma^{F}(0)\right) \\
\sigma^{F}(t) &=& (g_{f}^{F}(t))^{2} g_{f}^{F}(-t)
\end{eqnarray}
(as explained in  Appendix. \ref{AppendiceNum}, a sign change occurs in the 
expression of the self-energy at $T=0$)

\begin{figure}
\[
\fig{10cm}{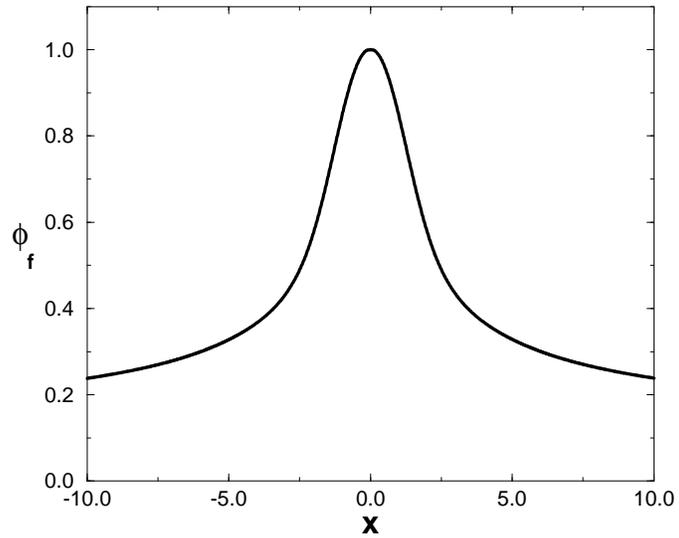} 
\]
\vskip 1cm
\caption{ \label{FntEchelleTnulDopage}
$T=0$ scaling function associated with the spectral density
$\rho_{f}(\omega ) = \frac{1}{t\delta } \phi_{f}(\omega /\epstar)$
in the low-doping regime.}
\end{figure}

The universal scaling function $\phi_f$ can be obtained by solving 
numerically Eq. (\ref{Equationgf}), and the result is displayed in 
Fig. \ref{FntEchelleTnulDopage}.
The asymptotic behaviours of $\phi_{f}(\overline{\omega})$ for large 
and small $\overline{\omega}=\omega/\epstar$ can be obtained analytically and 
read: 
\begin{eqnarray}
\nonumber
\phi_{f}(\ombar) &= \frac{1}{\pi } - c_{1} \ombar^{2} +\cdots 
\hskip 2cm &\hbox{for } \ombar\rightarrow 0 \\
\phi_{f}(\ombar)&= \frac{\displaystyle c_{2}}{\displaystyle \sqrt{\ombar}}
+\cdots
\hskip 2cm &\hbox{for } \ombar\rightarrow +\infty  
\end{eqnarray}
where $c_{1}$ and $c_{2 }$ are two constants. 
The low-frequency behaviour reflects the Fermi-liquid nature of the 
low-energy excitation spectrum, while the $1/\sqrt{\omega}$ behaviour 
characteristic of the undoped spin-liquid is recovered for 
$\omega>\epstar$.

\subsection{Single electron properties at $T=0$ :
 quasiparticle residue, effective mass, Luttinger theorem and photoemission}
\label{PropUnePart}

In this section, we focus on the one-particle Green's function for the 
physical electron, which is related to that of the auxiliary fermion by: 
\begin{equation}\label{DefGc}
G_c(\k,i\omega_n) = - <T c_{\k\alpha} c^\dagger_{\k\alpha} > 
= - <T b^\dagger_{\k}f_{\k\alpha} 
    f^\dagger_{\k\alpha}b_{\k} > = 
    {{M\delta}\over{2}} G_f(\k,i\omega_n)  
\end{equation}
hence:
\begin{equation}\label{greenphysique}
{{M\delta}\over{2}}G_c(\k,i\omega_n)^{-1}= 
i\omega_n+\mu-\lambda_0-\Sigma_f(i\omega_n)-\delta\epsilon_{\k}
\end{equation}
In this expression, $\epsilon_{\k}$ stands for the one-particle energies of 
a non-interacting tight-binding model on the Bethe lattice with hopping 
$t/\sqrt{z}$ between nearest neighbour sites \cite{footnoteFourier}.
The distribution of these single-particle energies is the semi-circular 
density of states $D(\epsilon)$ defined in (\ref{DefD}).

From the large-frequency behaviour of Eq.(\ref{greenphysique}), we see 
that the physical electron spectral density in the $M\rightarrow\infty$ limit 
is normalized as $\int_{-\infty}^{+\infty}\rho_c =M\delta/2$ (in contrast 
$\int_{-\infty}^{+\infty}\rho_f=1$). This is expected from the fact that 
the constraint (\ref{DefConstraint}) on the Hilbert space yields a
 normalisation (for arbitrary $M$)  
$\int_{-\infty}^{+\infty}\rho_c = <\{c,c^\dagger\}>=M\delta/2+(1-\delta)/2$ 
(note that this yields $(1+\delta)/2$ for $M=2$, as expected for the 
$U=\infty$ Hubbard model). 

Since our normalisation of the hopping is $t_{ij}=2t/(M\sqrt{z})$, the 
 non-interacting conduction electron Green's function reads:
\begin{equation}
G_c(\k,i\omega_n)_{free}^{-1}=i\omega_n+\mu-{{2}\over{M}}\,\epsilon_{\k}
\end{equation}
Thus, the physical electron self-energy reads:
\begin{equation}
\label{Sigmac}
\Sigma_{c}(i\omega_n)= 
i\omega_n +\mu - 
\frac{2}{M\delta }
\left( i\omega_n +\mu -\lambda_{0}-\Sigma_{f}(i\omega_n) \right)
\end{equation}
We observe that it depends solely on frequency, as is generally the case 
in the limit of large dimensionality \cite{review}. 

We first consider the location of the Fermi surface for both the 
non-interacting and interacting problems, {\it i.e} look for the poles of the 
electron Green's function. 
In the non-interacting case, we relate the chemical potential at $T=0$ 
to the number 
of particles $<n_{\alpha}>=(1-\delta)/2$ and find:
\begin{equation}
\mu_{free} = {{2}\over{M}}\, \mu_0(\delta)
\end{equation}
where the function $\mu_0(\delta)$ is defined by the relation:
\begin{equation}
\label{DefMu0}
\int_{-\infty }^{\mu_{0}(\delta)} d\epsilon \,D(\epsilon ) =
\frac{1-\delta }{2}
\end{equation}
Hence the non-interacting Fermi surface corresponding to a doping $\delta$ 
is defined by $\epsilon_{\k}=\mu_0(\delta)$. 
In the interacting case, we see from Eq.(\ref{greenphysique}) that  
 the Fermi surface is located at  
$\epsilon_{\k}=\left(\mu(T=0)-\lambda_0(T=0)-
\Sigma_f(\omega=0,T=0)\right)/\delta$. In the absence of magnetic scattering 
($J=0$), it can be shown by an explicit calculation \cite{largen} 
from the saddle-point 
equations that the r.h.s of this equation is just $\mu_0(\delta)$ and thus 
that the Fermi surface is unchanged in the presence of the constraint. 
When $J\neq 0$, such an explicit calculation is not possible, since the 
saddle point equations are coupled non-linear integral equations. 
However, a proof of 
Luttinger theorem can still be given  using the fact that a Luttinger-Ward 
functional exists for this problem and is known in explicit form in the 
large-$M$ limit, as detailed in Appendix \ref{Lutt}.
The conclusion of this analysis is that the volume of the Fermi surface
corresponds to $(1-\delta)/2$ particles per spin flavour and that the zero
temperature, zero frequency  self energy must obey :
\begin{equation}\label{luttFinal}
\mu(T=0)-\lambda_0(T=0)-
\Sigma_f(\omega=0,T=0) = \delta\mu_0(\delta)
\end{equation}
  
We now consider the weight and dispersion of the quasiparticles, that 
can be read off from Eqs.(\ref{greenphysique},\ref{Sigmac}) by expanding 
around the Fermi surface. We define a renormalisation factor for the 
auxiliary fermions as:
\begin{equation}\label{DefZ}
Z_{f}= \left.
\left(1-{{\partial\Sigma_f}\over{\partial\omega}}\right)^{-1}
\right|_{\omega =0}
\end{equation}
so that the physical electron quasiparticle residue reads:
\begin{equation}\label{RelationZcZf}
Z_{c}= \frac{M}{2}\,\delta Z_{f}
\end{equation}
From the low-frequency analysis of the preceding section and the 
corresponding estimates of the coherence scale, we expect $Z_{c}$ to be of
order $\epstar /t$ and thus :
\begin{equation}
Z_c \sim {{t}\over{J}}\,\delta^2 \sim 
\frac{\delta^{2} }{\delta^{*}}
\,\,\, (\delta\ll\delta^*)\,\,\,,\,\,\,
Z_c \sim \delta \,\,\, (\delta\gg\delta^*) 
\end{equation} 

\begin{figure}
\[
\fig{10cm}{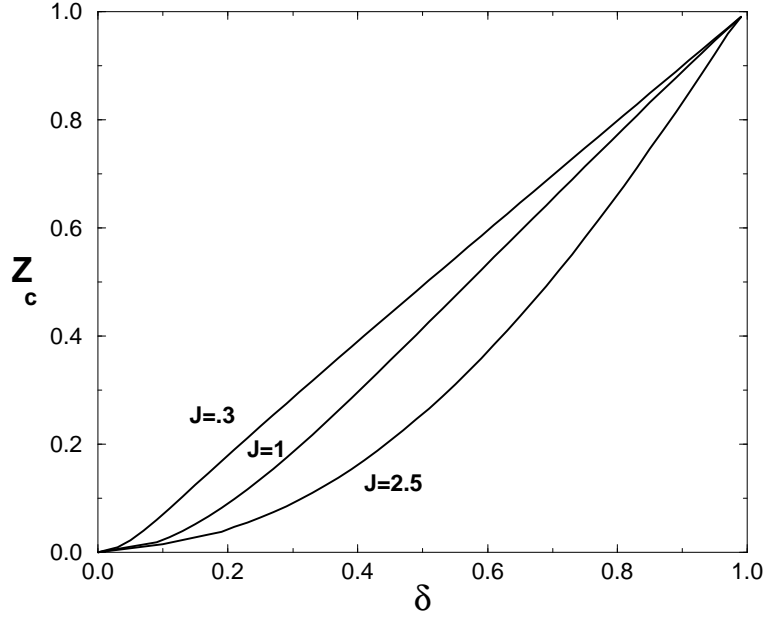} 
\]
\vskip 1cm
\caption{ \label{ZfntDelta}
Physical electron quasi-particle residue $Z_{c}$ vs. doping for $J=0.3,1,2.5$ 
(the proportionality factor $2/M$ has been set equal to $1$)}
\end{figure}

In Fig.\ref{ZfntDelta}, we display the result of a numerical calculation 
of $Z_{c}$ as a function of doping, for three values of 
$J/t$. These results entirely confirm the above  expectations. 
We have checked that at small doping $Z_{c}/\delta^{*}$ scales proportionally
to $(\delta /\delta^{*})^{2}$ with a universal prefactor.

From Eq.(\ref{greenphysique}), we see that the quasiparticles have a 
dispersion characterized by an effective hopping $t_{eff}/t=\delta Z_f$ 
($m^*/m = 1/Z_c \propto 1/(\delta Z_f)$). Hence the effective mass diverges 
as the Mott insulator is reached (as $1/\delta^2$). 
The reason for this divergence is the large (extensive
\cite{footentropy}) entropy of the insulating spin-liquid
ground-state. This entropy must be released at a temperature of the
order of the coherence scale $\epstar$ in the doped system. Hence,
integrating the specific heat ratio $C/T=\gamma$ between $T=0$ and
$T=\epstar$ leads to $\gamma\epstar\sim 1$, which is the result found
above. This divergence of $\gamma$ as $\delta\rightarrow 0$ is clearly
an artefact of the large-$M$ and large-$d$ limit. The residual
ground-state entropy of the spin-liquid phase should not survive a
more realistic treatment of this phase (whether this happens while
preserving $\chi''(\omega)\sim\mbox{const.}$ in this phase is an open
problem at this moment). Furthermore, our model does not include a 
uniform antiferromagnetic exchange constant superimposed on the random
part. Including this coupling will help locking the spins into
singlets and cutoff the divergence of the effective mass (for a
large-$M$ treatment of this point, see e.g Ref.\cite{largen}).

\begin{figure}
\[
\fig{12cm}{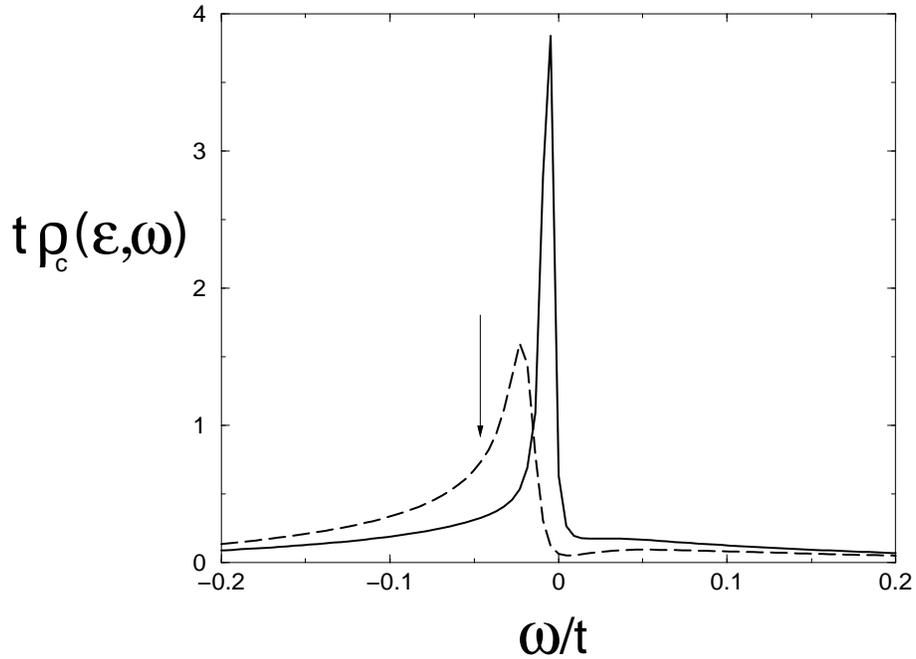} 
\]
\vskip 1cm
\caption{ \label{FigPhoto}
Conduction electron spectral density $\rho(\epsilon_{\k},\omega)$ for  
 $\delta =0.04$, $T/t = 1/300$ and $J/t=0.3$ and for 2
 values of the energy $\epsilon_{\k}$. The arrow indicates the cross-over
between the Fermi-liquid regime and the spin-liquid regime, as explained in the
text. 
}
\end{figure}

Finally, we discuss the shape of the conduction electron
spectral density $\rho (\epsilon_{\k} ,\omega )$ for a {\sl fixed}  value of
the energy $\epsilon_{\k}$ as a function of frequency,  as relevant for
photoemission experiments : 
\begin{equation}\label{DefRhoc}
\rho_{c}(\epsilon_\k,\omega ) = -\frac{1}{\pi }G''_{c}(\epsilon_\k,\omega )
= -\frac{M \delta}{2\pi} \frac{\Sigma_{f}''(\omega)}
{\left(\omega +\mu -\lambda_{0}-\Sigma_{f}'(\omega) -
\delta\epsilon_\k\right)^{2}
 + \Sigma_{f}''(\omega )^{2}}
\end{equation}
Numerical results for this quantity are displayed on Fig. \ref{FigPhoto}.   
This function is peaked at a frequency $\omega_{peak} \simeq
Z_{c}(\epsilon_{\k}-\epsilon_{\k_{F}})$ with a height  
of order $1/\omega_{peak}^2$ (at $T=0$). Moving away from this quasiparticle 
peak, $\rho_{c}(\epsilon_\k,\omega )$ has the characteristic 
$1/\omega^2$ decay of a Fermi 
liquid only in the limited frequency range
$|\omega_{peak}|<|\omega|<\epstar$, followed 
(for $\delta<\delta^*$) by a much slower 
$1/\sqrt{\omega}$ tail corresponding to the spin liquid  
regime in the frequency range $\epstar<|\omega|<J$. (We note that this 
non-Fermi liquid tail is absent in the high-doping regime).  
These two regimes are clearly apparent on 
Fig. \ref{FigPhoto}. If the resolution of a photoemission experiment is 
not significantly smaller than $\epstar$, the peak will be smeared into a 
broad feature, and the measured signal will be dominated by the slowly 
decaying tail. Furthermore, as shown in the next section, temperature has a 
large effect on the peak, whose height decreases as $1/\sqrt{T}$ in the 
temperature range $\epstar<T<J$.  

\subsection{Finite-temperature crossovers}
\label{SectCrossOverTherm}
The metal-insulator transition at $T=0$ as $\delta \rightarrow 0$ is
 a quantum critical point. The associated crossover regimes at finite
temperature can be  easily deduced by 
comparing the coherence scale $\epsilon_F^*$ to the magnetic exchange $J$ 
and to the temperature. This analysis yields three regimes, as depicted on 
Fig.\ref{Crossovers}:

\begin{itemize}

\item For $T<\epsilon_F^*$, the doped holes form a Fermi liquid. The 
low-energy degrees of freedom are the fermionic quasiparticles described 
by the auxiliary fermions $f_{\alpha}$, which behave in a coherent
 manner since 
their inverse lifetime vanishes at low-frequency as 
$\Im\Sigma_f\propto\omega^2$ in this regime.

\item At low doping $\delta<\delta^*$, an intermediate 
temperature regime exists, defined by: $\epstar<T<J$. In this regime, 
coherent quasiparticles no longer exist (as shown below, 
$\Im\Sigma_f\propto\sqrt{\omega}$), but the spin degrees of freedom are not 
free local moments since the temperature is smaller than the 
magnetic exchange. Hence, the spins behave in this regime as in a spin liquid, 
with a {\it marginal Fermi liquid 
form for the local spin response function}. As shown below, this regime
corresponds to the so-called ``quantum-critical'' regime associated with the 
quantum critical point at $T=\delta =0$.
In this regime, the
low-energy scale $\epstar $ drops out from response functions, which obey 
 universal scaling properties as a function of the ratio $\omega /T$.

\item Finally, a high-temperature regime applies, defined by 
$T>J$ (for $\delta<\delta^*$) or $T>\delta t=\epsilon_F^*$ 
(for $\delta>\delta^*$) in which both spin and charge are incoherent and 
essentially free. We note that if $J<t$ and the doping is larger than 
$\delta^*\simeq J/t$, the system goes directly from a Fermi liquid 
to this high-temperature regime as temperature is increased, without an 
intermediate marginal Fermi liquid regime. 

\end{itemize}

This qualitative analysis can be established on firmer grounds by 
generalizing the low-doping scaling analysis of Sec.\ref{LowFreqAnalysis}
 to finite temperature.
 Assuming that the coherence scale is small as compared 
to both $J$ and the hopping $t$ (i.e that $\delta<\delta^*$), and 
that $\omega, T \ll J,t$,  the spectral function takes 
the following scaling form, generalizing Eq.(\ref{DefPhif}):
\begin{equation}\label{rhoscalefiniteT}
\rho_f(\omega,T) = {{1}\over{t\delta}} 
\Phi_{f}\left({{\omega}\over{\epsilon_F^*}},
{{T}\over{\epsilon_F^*}}\right) 
\end{equation}
In the following, $\ombar$ and $\Tbar$ stand for $\omega/\epstar$ and 
$T/\epstar$, respectively.  
Eq.(\ref{rhoscalefiniteT}) yields for the Green's function
 $G_{f}(\omega ) = \frac{1}{t\delta} g_{f}(\ombar,\Tbar)$  
(with $\Phi_{f}=-g_{f}''/\pi$). We assume that the self energy
also scales  as
$\Sigma_{f}''(\omega ) = (t\delta ) \sigma_{f}''(\ombar,\Tbar)$.
From  the saddle-point equation (\ref{EqBase}) we deduce
 the following equations :

\begin{eqnarray}\label{FullScaling}
\Im g_{f}^{-1}(\ombar,\Tbar) &=& - g_{f}''(\ombar,\Tbar) - 
\sigma_{f}''(\ombar ,\Tbar ) \\
\nonumber
\sigma''_{f}(\ombar,\Tbar) &=& \pi \int_{-\infty }^{\infty }
\!\int_{-\infty }^{\infty } d x_{1} d x_{2}\,\,
\Phi_{f}(x_{1},\Tbar)\Phi_{f}(x_{2},\Tbar)
\Phi_{f}(x_{1}+ x_{2} -\ombar ,\Tbar)\\
&& 
\left(
n_{F}\left(\frac{x_{2}}{\Tbar} \right) - 
n_{F}\left(\frac{x_{1}+x_{2}-\ombar }{\Tbar} \right)
\right)
\left(
n_{F}\left(\frac{x_{1}}{\Tbar} \right) + 
n_{B}\left(\frac{x_{1}-\ombar }{\Tbar} \right)
\right)
\end{eqnarray}
 In this expression, $n_{F}$ and $n_{B}$
are the Fermi and Bose factor :
$n_{F,B}(y) = 1/(e^{y} \pm 1)$.
With a  Kramers-Kronig transformation one can deduce 
$g_{f}'$ and $\sigma_{f}'(\ombar,\Tbar )-\sigma_{f}'(0,\Tbar)$ 
from  $\Phi_{f}$ and $\sigma_{f}''$.
From the equation for $\Re g_{f}^{-1}$, we have that 
$\mu -\lambda_{0} - \Sigma_{f}'(\omega =0,T) \sim f(T/\epstar)$
where $f$ is some scaling function which can in principle
be calculated from (\ref{FullScaling}). The function $f$ vanishes at small
argument ($f(0)=0$) due to Luttinger's theorem,  and at large argument
($f(+\infty ) = 0$) due to the auxiliary fermion particle hole symmetry 
of the undoped model.

We now discuss the solution of this scaled integral equation, 
and the form taken by $\Phi_f$ in the various regimes described above. 

i) {\it Fermi liquid regime} :  $T\ll\epstar$. 
At zero temperature,  
the scaling function $\Phi_f$ reduces to that in Eq.(\ref{DefPhif}):
\begin{equation}
\Phi_f(\ombar,\Tbar=0)=\phi_f(\ombar)
\end{equation}
We can also consider the limit of low-frequency and temperature 
$\omega,T\ll\epstar$ but with an arbitrary ratio $\omega/T$. In this limit, 
the self-energy term is negligible altogether in Eq.(\ref{FullScaling}), and 
one gets simply $g_{f}^2=-1$ i.e:
\begin{equation}\label{PhiLF}
\Phi_f(\ombar\ll 1,\overline{T}\ll 1)\rightarrow {{1}\over{\pi}}
\end{equation}
Note that the r.h.s could a priori be a function of the ratio 
$\omega/T$, but is actually a constant (as is generically the 
case in a Fermi liquid). From this, we can deduce a scaling form of the 
scattering rate in the same regime. Indeed (\ref{PhiLF}) corresponds to the 
imaginary time Green's function:
\begin{equation}
G_f(\tau) \rightarrow - {{1}\over{\pi t\delta}} 
{{\pi/\beta}\over{\sin\pi\tau/\beta}}\qquad \qquad 
1/\epsilon_F^*\ll\tau\,,\,\beta-\tau
\end{equation}
Hence, in this limit, the self-energy takes the form:
\begin{equation}
\Sigma_f(\tau) \sim  - {{J^2}\over{(\pi t \delta)^3}}\,
\left({{\pi/\beta}\over{\sin\pi\tau/\beta}}\right)^3
\end{equation}
which can be Fourier transformed to yield:
\begin{equation}\label{ScattFL}
\Im\Sigma_f(\omega\ll\epsilon_F^*,T\ll\epsilon_F^*) = 
- {{J^2}\over{2 (\pi t \delta)^3}}(\omega^2+\pi^2 T^2)
\end{equation}

ii) {\it Spin-liquid regime :} $T\gg\epstar$.  
In this quantum critical regime 
the energy scale $\epstar$ drops out from the problem 
and the spectral density and response functions become functions
of the ratio $\omega/T$ only. 
Indeed, the  scaling function $\Phi_f(\ombar,\Tbar)$ takes the form 
$\varphi_{f}(\ombar/\Tbar)/\sqrt{T}$ in the limit $\Tbar\gg  1$.
In order to find $\varphi_{f}$ in explicit form, 
we divide both side of (\ref{FullScaling}) by $\sqrt{\Tbar }$
and take the limit $\Tbar \rightarrow \infty$, $\ombar /\Tbar$ fixed.
Then the first term of the r.h.s vanishes and we
are left with a scaled equation for $\varphi_{f}$ in which all dependence on
$\epstar$ has disappeared.
Remarkably, this integral equation can be solved in closed form and yields :
\begin{equation}\label{ScalingFormRho}
\rho_f(\omega,T)\rightarrow 
\frac{1}{\sqrt{JT}}\,\, \varphi_{f}\left(\frac{\omega }{T} \right) = 
\frac{1}{2 \pi^{\frac{9}{4}}\sqrt{JT}}
\cosh \left(\frac{\omega }{2T} \right)
\left|\Gamma \left(\frac{1}{4} + i\frac{\omega }{2\pi T} \right) \right|^{2}
\end{equation}
Some details are provided in  Appendix \ref{AppFonctionsEchelle}.
This scaling function describes how
 the $1/\sqrt{\omega}$ singularity 
associated with the low-energy excitations of the 
spin-liquid is cutoff (by the temperature)  at frequencies 
$\omega<T$ so that the spectral density is of order $1/\sqrt{JT}$ 
 at $\omega=0$.   
(Note that if the limit  $\Tbar \rightarrow \infty $ is performed while keeping
$\ombar$ fixed in (\ref{FullScaling}), the same result is obtained as when 
the limit is taken with  $\ombar =0$. Hence there is no additional crossover
in the frequency dependence of the response functions below
 $\omega =\epstar$ in this regime).
Eq.(\ref{ScalingFormRho}) corresponds to the following 
scaling form for the imaginary time Green's function:
\begin{equation}\label{GEchelle}
G_f(\tau) \sim -
\frac{1}{\sqrt{2 J} \pi^{\frac{1}{4}}}  \,
\left({{\pi/\beta}\over{\sin\pi\tau/\beta}}\right)^{1/2}
\end{equation}
Remarkably, (\ref{GEchelle}) has the form which would hold in a model 
having {\it  conformal invariance}, for example a quantum impurity
model of a spin interacting with a structureless bath of conduction 
electrons. In that case, a conformal mapping from the $T=0$ half plane
$\tau >0$ to the finite-temperature strip $0\leq\tau\leq\beta$ can be
used to show \cite{Tsvelik} that if the Green's 
function decays as $1/\sqrt{t}$ at 
$T=0$, then it takes a scaling form given by (\ref{GEchelle}) at
finite temperature (lower than a high-energy cutoff). In the present
case, the original model is an infinite connectivity  {\sl lattice model}
 which does not {\it a priori} satisfy conformal invariance. It does map onto a
single-site quantum impurity model, but with  an
additional self-consistency condition.
This means that the effective bath for the local 
spin is given by the local spin-spin correlator itself, and thus does 
have non-trivial structure at low-energy. However, this structure
appears only as a subdominant correction to the leading low-frequency
behaviour $\chi''_{loc}(\omega)\sim\mbox{const}$. For this reason,
our effective single-site model does obey conformal invariance
properties in the low-energy limit, which explains the result
above. This remark actually applies in a broader context than the
specific model considered here, as will be discussed in more details
elsewhere.   

Let us also consider the scattering rate in this regime, which is 
obtained by Fourier transforming the imaginary time self-energy:
$ \Sigma_f(\tau) \sim   - 
\sqrt{J}
\left({{\pi/\beta}\over{\sin\pi\tau/\beta}}\right)^{3/2}
/ (4\pi)^{\frac{3}{4}}$
which yields:
\begin{equation}\label{ScalingFormSigma}
\Sigma_f''(\omega) \sim -\pi^{-\frac{3}{4}}\sqrt{JT}
\cosh \left(\frac{\omega }{2T} \right)
\left|\Gamma \left(\frac{3}{4} + i\frac{\omega }{2\pi T} \right) \right|^{2}
\end{equation}

We have calculated numerically the real-frequency, 
finite-temperature Green's function 
by following the method described in Appendix \ref{AppendiceNum}. 
On Fig. \ref{DosFntTemp}, we display results for the spectral density 
for various temperatures for $J/t = 0.3$ at a doping of 
$\delta= 0.04 < \delta^{*}$.
These values correspond to a low energy coherence scale 
$\epstar /J = (\delta /\delta^{*})^{2} \simeq 1.8 \, 10^{-2}$. 
The crossover from the Fermi-liquid regime at low temperature 
into the quantum critical regime at intermediate temperatures is clearly
visible (in particular, the peak height can be checked to decrease as 
$1/\sqrt{T}$). Note also that  $\rho_{f}(\omega )$ remains approximately
centered at $\omega \approx 0$ until $T\simeq J$ and shifts rapidly away from
$\omega =0$ for $T > J$. In the inset, we also display the thermal scaling
function associated with $\rho_{f}$, Eq. (\ref{ScalingFormRho}).   

\subsection{Local spin dynamics}

\begin{figure}
\[
\fig{10cm}{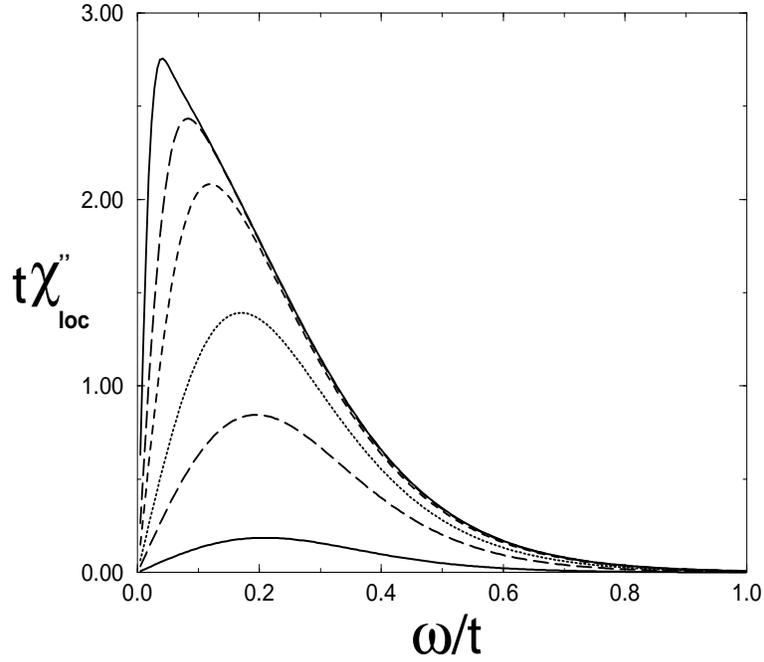} 
\]
\vskip 1cm
\caption{ \label{KilocFntT}
Local dynamical susceptibility $\chiloc''(\omega ) $ for $\delta =0.04$ and
$J/t=0.3$. The different curves
correspond from top to bottom to $T/t = 1/200,1/50,1/25,1/10,1/5,1$. In the
temperature range $\epstar <T<J$ and for frequencies $\omega <J$, these curves
scale on the universal form  
$\chi_{loc}''(\omega,T) = \frac{\sqrt{\pi}}{2J}  \tanh \frac{\omega}{2T} $
}
\end{figure}

In this section, we describe the behaviour of the local spin dynamics 
in the various temperature regimes. In the large-$M$ limit, the local 
spin correlation function is given by:
\begin{eqnarray}
\nonumber 
\chiloc(\tau) &=& - G_f(\tau)G_f(-\tau)  \\
\chiloc''(\omega) &=& \pi \int_{-\infty }^{+\infty } d \nu 
 \rho_f(\nu) \rho_f (\nu -\omega ) 
\left(n_{F}(\nu -\omega ) - n_{F}(\nu) \right)
\end{eqnarray}
In Fig. \ref{KilocFntT}, we display $\chiloc''(\omega )$ for various
temperatures and the same choice of parameters as in Fig. \ref{DosFntTemp}. 
In the low doping regime, $\epstar \ll J,t$, $\delta <\delta^{*}$, 
$\chiloc''$ obeys a scaling form which follows from the  convolution 
of (\ref{rhoscalefiniteT}) : 
\begin{equation}\label{ScalFormGenKi}
\chiloc''(\omega,T) = \frac{1}{J}\,
\Phi_{\chi}
\left(\frac{\omega }{\epstar},\frac{T}{\epstar} \right)
\end{equation}
Let us  discuss  the limiting forms of this expression as $T\rightarrow 0$ 
or  $T\gg\epstar$.

\begin{itemize}
\item [i)] {\sl At zero temperature,} $\chiloc''(\omega)$ has a shape which
resembles the undoped spin-liquid case (Fig. \ref{DosSpinFluid}) for
frequencies $\omega >\epstar $. 
At lower frequency, the Fermi liquid behaviour $\chiloc''(\omega ) \propto
\omega $ is recovered. This results in a peak with a  height of order  
$1/J$ at $T=0$. 
This crossover can be described by a scaling function : 
\begin{equation}
\chiloc''(\omega\ll J,T=0)= {{1}\over{J}} \phi_{\chi}
\left(\frac{\omega }{\epstar} \right)
\end{equation}
where $\phi_{\chi}(x) = \Phi_{\chi}(x,y=0)$
 can be obtained by convoluting $\phi_{f}$ with itself
resulting in the asymptotic behaviours : 
\begin{eqnarray}\label{CrudeEstKi}
\nonumber
\chiloc''(\omega,T=0) &\simeq \frac{\omega}{\pi (\delta t)^2} \,\,\,
&\omega\ll\epstar\\
&\simeq \frac{\pi ^{\frac{3}{2}}}{2J}  & \,\,\,\epstar\ll\omega\ll J
\end{eqnarray}
This can be used to estimate the behaviour of the static local susceptibility
 at low doping : 
$\chi'_{loc}(\omega=0) = \int d\omega \chi''_{loc}(\omega) /\omega$. 
In this integral, the region $\epstar <\omega < J$ (corresponding to spin
liquid excitations) gives the dominant contribution, leading to the
logarithmic behaviour for $\delta \ll \delta^{*}$: 
\begin{equation}\label{EstimateKiloc}
\chi'_{loc}(\omega=0) \simeq \frac{1}{J}
 \ln {{\delta}\over{\delta^*}}
\end{equation}
In contrast, as detailed in Appendix \ref{Susceptibilite},  
the {\it uniform } static susceptibility 
$\chi = \chi'(\q=0, \omega =0)$  is a constant of order $1/J$, with no
divergence at small doping. 

\item [ii)] In the {\sl  spin-liquid regime } $T\gg\epstar$,  
$\chi_{loc}''$ becomes a function of $\omega/T$.
The corresponding scaling function is remarkably simple : 
from Eq. (\ref{GEchelle}) we have 
$\chi_{loc}(\tau) \propto \pi /(\beta \sin (\pi \tau /\beta ))$
which yields :
\begin{equation}\label{MflChi}
\chi_{loc}''(\omega,T) = \frac{\sqrt{\pi}}{2J}\tanh \frac{\omega}{2T}
\end{equation}
This behaves exactly as the spin response function postulated in 
the marginal Fermi liquid phenomenology \cite{mfl}
 ($\omega/T$ for $\omega<<T$, $const.$ for $\omega>T$). 
\end{itemize}

\begin{figure}
\[
\fig{10cm}{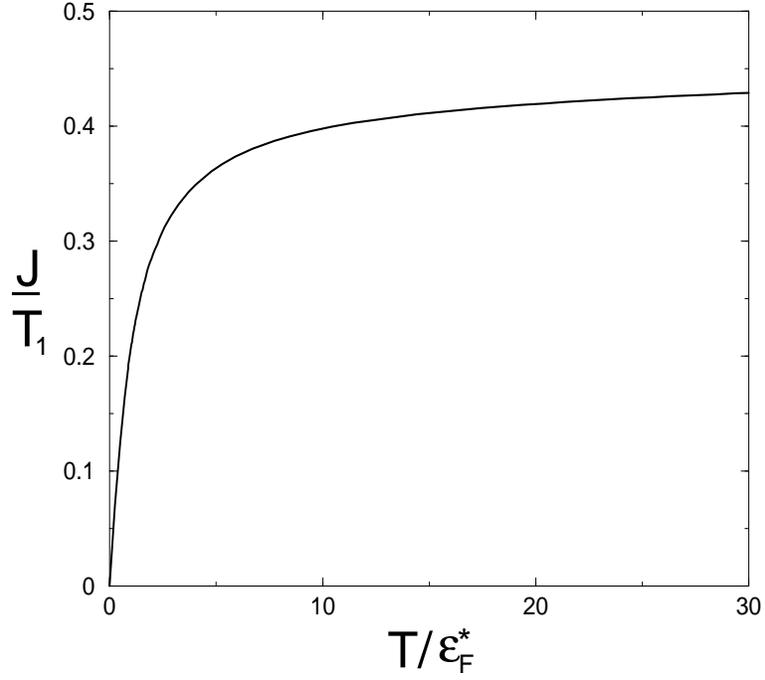} 
\]
\vskip 1cm
\caption{ \label{NMRrate}
Scaling function $\psi$ associated with the NMR relaxation rate : $J/T_{1} =
\psi (T/\epstar )$}
\end{figure}

We finally use these results to compute the temperature dependence of the
 NMR relaxation rate : 
\begin{equation}
\frac{1}{T_1T} = 
\left.\frac{\chi''_{loc}(\omega,T)}{\omega} \right|_{\omega=0}
\end{equation}
Expanding the scaling form (\ref{ScalFormGenKi}) to linear order in $\omega$ 
(and noting that $\Phi_{\chi}(0,y)=0$ because $\chi''$ is odd), we 
get for $T\ll J$:
\begin{equation}
\frac{1}{T_1} = \frac{1}{J} 
\psi\left(\frac{T}{\epstar} \right)
\end{equation}
(with $\psi(y)=y \partial_x\Phi_{\chi}(x=0,y)$). 
In Fig. \ref{NMRrate}, we plot this universal scaling function. We have also
checked the data collapse of our numerical results  on this function.
Limiting forms are easily obtained from Eq. (\ref{CrudeEstKi}) and
(\ref{MflChi})  :
\begin{itemize}
\item [i)] 
$T\ll\epstar$: $\psi(y\ll 1)\sim y/\pi $ 
Hence $\frac{1}{T_1} \simeq {T \over \pi  (\delta t)^2}$.
 We find a Korringa 
law (as expected from a Fermi liquid) but with a {\it very strong doping 
dependence}. We also note that in contrast to a non interacting Fermi gas,
$1/(T_{1}T) \propto 1/(t\delta)^{2}$, $\chiloc \propto 1/J \ln
(\delta^{*}/\delta)$  and $\chi \propto 1/J$ obey quite different
behaviour as a function of doping. In particular the so-called ``Korringa
ratio'' $1/(T_{1}T \chi^{2})\simeq (\delta^{*} /\delta)^{2}$  is  very
large at low doping. 
 \item [ii)]
$J>T\gg\epstar$: $\psi(y\gg 1)\rightarrow \sqrt{\pi}/4$, 
hence $\frac{1}{T_{1}} \sim \frac{\sqrt{\pi}}{4J} = \const$ 
as expected in a {\it marginal 
Fermi liquid}. We note that $1/T_{1}$ is doping-independent in this
quantum-critical regime.
This is because the scale $\epstar $ no longer appears explicitly.
\end{itemize}

\subsection{Transport and frequency-dependent conductivity}

In the limit of large connectivity, the current-current correlation function
has no vertex corrections, due to the odd parity of the current (See
e.g. \cite{review}). Hence the frequency dependent conductivity is given by :
\begin{equation}\label{ConducBase}
\Re \sigma(\omega ) =t^{2} \intR  d \epsilon  D(\epsilon ) \intR d \nu 
\rho_{c}(\epsilon ,\nu ) \rho_{c}(\epsilon ,\nu +\omega )
\frac{n_{F}(\nu ) - n_{F}(\nu +\omega )}{\omega }
\end{equation}
where $\rho_c(\epsilon,\omega)$ is the single-electron spectral 
density defined in Eq.(\ref{DefRhoc}). This expression yields the conductivity
in units of $e^{2}/(h a^{2-d}) $ where $a$ is the lattice spacing and some
numerical prefactors have been dropped (we shall also ignore 
the prefactor $M$ in $\rho_c$).   

\subsubsection{Resistivity}
We first discuss the behaviour of the dc-conductivity :
\begin{equation}\label{ConducStat}
\sigma_{dc}(T) = \Re \sigma(\omega =0,T) 
 = \intR d\epsilon D(\epsilon) \intR \frac{dx}{4 \cosh^{2}
\left(\frac{x}{2} \right)} \,\,\,\,
\rho_{c}^{2}(\epsilon,Tx)
\end{equation}
 
{\sl i) In the Fermi liquid regime } $T\ll \epstar $, we have from the
behaviour (\ref{ScattFL}) of the scattering rate : 
 $-\mbox{Im}\Sigma_f(\omega,T)\propto J^2(\omega^{2} + 
\pi^{2} T^{2})/(\delta t)^3$ and 
$\omega+\mu-\mbox{Re}\Sigma_f(\omega,T) = \omega /Z_{f}  + \const T $.
 Making the change of variables $\epsilon=T u$,
 we see that the integral over $u$ in 
$\sigma_{dc}/T$ diverges as $1/T^{3}$. Hence, we find in this regime the  
expected Fermi-liquid behaviour of the resistivity $\rho_{dc}=1/\sigma_{dc}$:
\begin{equation}\label{ResisLiqFermi}
\rho_{dc}(T) \propto \left( \frac{T}{\epstar}\right) ^{2}
  \qquad \qquad \qquad  T\ll \epstar 
\end{equation}

{\sl ii) In the spin liquid regime } $\epstar\ll T\ll J $ at low doping,
$-\Sigma_f'(\omega ) $ 
is of order $\sqrt{JT}$ (times a scaling function of $\omega /T$).
This must be compared
to $\delta \epsilon \simeq \delta  t$ in the denominator of
$\rho_{c}(\epsilon,\omega)$. Since $T\gg \epstar $, we see that 
$\mbox{Re}\Sigma_f$ always
dominates over $\delta  \epsilon $ which can thus be neglected.
Hence one can replace  $\rho_{c}(\epsilon,\omega) $ by the {\sl local }
spectral function  $\delta \rho_{f}(\omega)$.
In other words the limit  $\delta \rightarrow 0 $ must be taken   before
 the low temperature limit in this quantum critical regime. 
Using the thermal scaling function Eq. (\ref{ScalingFormRho}), we obtain :
\begin{equation}\label{ConducStat2}
\sigma_{dc}(T)  =\frac{\delta^{2}}{16JT}\intR \frac{dx}{\cosh^{2}
\left(\frac{x}{2} \right)} \,\,\,\, 
\varphi_{f}(x)^{2}.
\end{equation}
The integral can calculated explicitly
using $\int_{0}^{\infty} d x \,
\left|\Gamma \left(\frac{1}{4} +i x \right) \right|^{4} = \pi^{3}$
(\cite{Grad}, Eq.  6.412),
we  finally find :
\begin{equation}\label{ResisSpinFluid}
\rho_{dc}(T)  =  16 \sqrt{\pi} \,\frac{T}{\epstar}  \qquad \qquad 
\epstar\ll T\ll J
\end{equation}
Hence the resistivity turns out to have a linear behaviour as a function of
temperature in the spin-liquid regime, again as in the marginal Fermi liquid
phenomenology. This is rather remarkable in view of
the fact that the {\sl single-particle} scattering rate  behaves as $\sqrt{T}$
in this regime. As further discussed in the conclusion, this is characteristic
of a regime of incoherent transport in which the  transport scattering
rate cannot be naively related to the single-particle lifetime. 
Furthermore, we note that the $\sqrt{\omega}$ behaviour of the self-energy 
is a crucial ingredient in producing a $T$-linear resistivity. With a 
different power law ($\omega^{\alpha}$), the resistivity would behave as 
$T^{2\alpha}$ in this incoherent regime.    

\begin{figure}
\[
\fig{10cm}{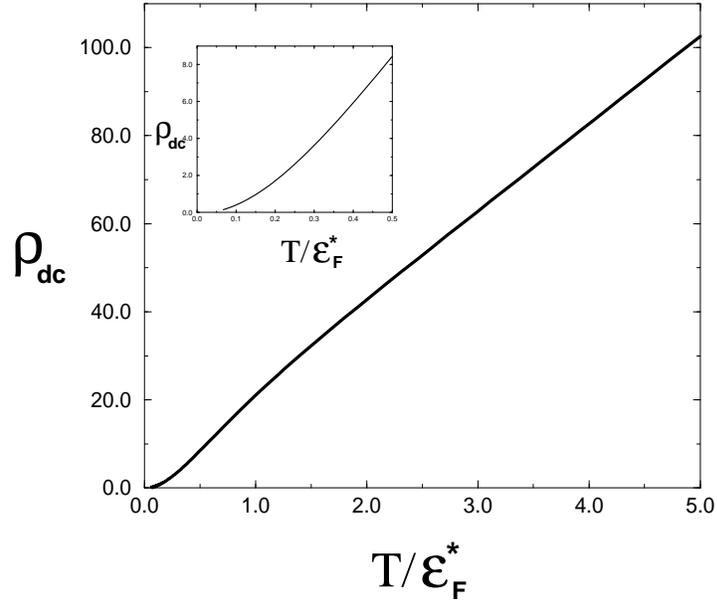} 
\]
\vskip 1cm
\caption{ \label{Resistivity}
Scaling function for the resistivity. Inset : low temperature Fermi liquid 
regime}
\end{figure}

The crossover from $T^{2}$ to $T$ in the resistivity can be captured in a more
precise manner in a universal scaling function :
\begin{equation}\label{ScalingResis}
\rho_{dc}(T) = \psi_{\rho}\left(\frac{T}{\epstar}  \right)
\end{equation}
We have determined numerically the function $\psi_{\rho }$, which is depicted
in Fig. \ref{Resistivity}.
We observe that  it is linear over a wide temperature range (with a slope in
agreement with (\ref{ResisSpinFluid})).

\subsubsection{Optical conductivity}
We now turn to the analysis of the frequency-dependent conductivity.

i) {\it In the Fermi liquid regime,}  
the conductivity takes  the form at $T=0$ :
\begin{equation}\label{ResigLiqFermi} 
\sigma (\omega ) = D \delta (\omega ) + \sigma_{reg}(\omega )
\end{equation}
where $D$ is the weight of the Drude peak and
 $\sigma_{reg}(\omega ) \rightarrow \const $ as $\omega \rightarrow 0$.
The Drude peak is easier to capture
 by a finite temperature analysis : the delta function  is
regularised by $T$ in the form $T^{2}/(\omega ^{2} + T^{4})$. Performing a
low-temperature, low-frequency analysis of (\ref{ConducBase}) leads to 
the estimation $D \propto  t^{2}D(\mu_{0}) Z_{f} \delta \propto \delta^{2}$
at small doping.
A closed formula can be given for $\Re \sigma_{reg}(\omega )$ as (truncated)
convolution of the scaling function $\phi_{f}$. A low-frequency analysis 
then shows that $\Re \sigma(\omega \ll \epstar)  =\const$, while 
$\Re \sigma(\omega \gg  \epstar) \sim \epstar /\omega   $.

{\sl ii) In the regime $\epstar <T<J$}, we have from (\ref{ConducBase}) and 
(\ref{ScalingFormRho}) the scaling form :
\begin{eqnarray}\label{DEfvarphisig}
\nonumber
\Re \sigma (\omega ) &=& \frac{\epstar}{\omega}\varphi_{\sigma} 
\left(\frac{\omega }{T} \right) \\
\varphi_{\sigma} (y) &\equiv& 
 \int_{-\infty }^{+\infty}\frac{dx}{\sqrt{|x(1+x)|}} \,\,
\varphi_{f} \left(x y \right)
\varphi_{f} \left((1+x) y \right)
\left[
f\left(x y \right) - 
f\left((1+x) y \right)
\right]
\end{eqnarray}
where $f(x)= 1/(e^{x} +1)$.
From (\ref{DEfvarphisig}),  $\varphi_{\sigma}(+\infty ) = \const$ and thus we
have in this spin-liquid regime  :
\begin{eqnarray}
\nonumber
\Re \sigma(\omega )   &\propto& \frac{\epstar}{\omega } 
\qquad \qquad \qquad \qquad 
T \ll \omega \ll J \\
\Re \sigma(\omega ) &\propto& \frac{\epstar}{T } 
\qquad \qquad \qquad \qquad 
 \omega \ll T 
\end{eqnarray}
Moreover, using the Kramers Kronig relation
 $\Im \sigma (\omega ) = \int d \omega' \Re \sigma (\omega')
 /(\omega -\omega')$ we find in the same regime for $\omega >T$ : 
\begin{equation}
\Im \sigma (\omega ) \propto \frac{\epstar }{\omega }
\ln \left(\frac{\omega }{\epstar } \right)
\end{equation}

Hence defining an optical scattering rate from an effective Drude formula :
$\tau_{opt}^{-1}(\omega) = \omega \Re \sigma (\omega) /\Im \sigma (\omega)$
we find $\tau_{opt}^{-1}(\omega) \sim \omega/ \ln (\omega/\epstar)$.

\begin{figure}
\[
\fig{10cm}{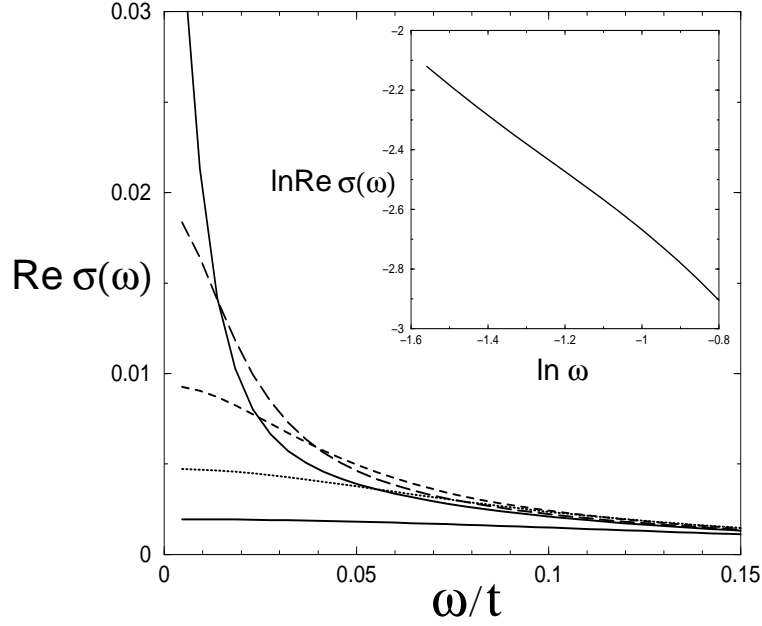} 
\]
\vskip 1cm
\caption{ \label{ReSigOm}
Real part of the optical conductivity $\Re \sigma (\omega )$ versus $\omega $,
for $\delta =0.04$ and $J/t=0.3$. The different curves correspond 
to $T/t = 1/200,1/100,1/50,1/25,1/10$.
Here $\epstar/t =1.8\,  10^{-2}$. 
Inset : the curve corresponding to $T/t = 1/100$,
plotted in log-log coordinates, in the frequency range 
 $T \sim \epstar <\omega < J$.
The $1/\omega $ behaviour described in the text is clearly visible.}
\end{figure}

We have also calculated $\Re \sigma(\omega) $ numerically, following the 
method explained in Appendix \ref{AppendiceNum}.
Numerical results are  displayed for various temperatures
on Fig.\ref{ReSigOm} and are in
agreement with the previous analysis. 

\section{Conclusion and discussion}

\subsection{Summary}

In this paper, we have solved a model of a doped spin-fluid with 
strong frustration on the exchange constants $J_{ij}$. 
The undoped model is an $SU(M)$ quantum Heisenberg 
model with random exchange, previously studied by Sachdev and Ye 
\cite{spinfluid} in the limit of large-$M$ and infinite connectivity. 
These authors found that, in this limit, quantum fluctuations are so strong 
that no spin glass phase forms \cite{warningbosons}.
Instead, a gapless spin liquid is found 
with local spin dynamics identical to the marginal Fermi liquid 
phenomenology \cite{mfl}. We generalised this result to finite 
temperature and found that the local spin response function displays 
$\omega/T$ scaling: $J\chi''(\omega,T)_{loc}\propto \tanh \omega/2T$ 
(for $\omega,T < J$). 
Doping this Mott insulating phase with holes, we found that a characteristic 
doping $\delta^*\simeq J/t$ appears separating two quite different doping 
regimes. In the high-doping regime $\delta>\delta^*$, magnetic effects 
are weak and a Brinkman-Rice Fermi-liquid description is valid, with a 
rather large coherence scale of order $\delta t$. In the low doping regime 
however, the interplay between local coherence and magnetic effects 
gives rise to a coherence scale $\epstar=J (\delta/\delta^*)^2$, which 
can be very low. At low temperature $T<\epsilon_F^*$, 
Fermi liquid behaviour is recovered, but an incoherent regime is found 
in a rather wide regime of temperature $\epsilon_F^*<T<J$ in which 
physical properties strongly deviate from Fermi liquid theory. This regime 
corresponds to the ``quantum critical regime'' associated with 
the metal-insulator transition which in this model happens at $\delta_c=T=0$. 
We found that both transport properties and response functions in this 
incoherent regime behave as in the  marginal Fermi liquid phenomenology, 
namely $\rho_{dc}\propto T$,   
$\tau_{opt}(\omega)^{-1}\propto\omega/\ln(\omega/\epstar)$,
 $1/T_1\propto\mbox{const.}$, and
$J\chiloc''(\omega,T)\propto \tanh\omega/2T$. 
Remarkably, 
{\it single-particle} properties deviate much more strongly from Fermi 
liquid theory, with a single-particle scattering rate behaving as 
$\Im\Sigma\propto\sqrt{\omega}$ (or $\sqrt{T}$), in contrast to the 
$\Im\Sigma\propto\omega$ behaviour postulated in the Marginal Fermi Liquid 
phenomenology.

These behaviour result from the solution of the large-$M$ saddle point 
equations, which also yields explicit expressions for the scaling functions 
of $\omega/\epstar$ and $\omega/T$ describing the crossover of the various 
physical quantities between the Fermi liquid and the non-Fermi liquid 
regime. We also note that in the large-$M$ limit, response functions can 
be calculated from the {\it interacting single-particle} Green's function. 
Hence  the  behaviour of $\Sigma\propto \sqrt{\omega }$ and of $\Im \chiloc
\propto  \const $ are intimately related.
In contrast, in the marginal Fermi liquid phenomenology, the behaviour of $\Im
\chi$ is related to a priori unknown higher order vertex functions.
In this sense, the present model yields a solution to the problem of
internal consistency of the Marginal Fermi Liquid Ansatz, resulting in a
more singular form of the single-particle Green's function.

We also note that numerical studies of the  doped two-dimensional 
$t-J$ model with uniform antiferromagnetic $J$ by Imada and coworkers
\cite{Imada} have some intriguing similarities with the results of the present
work. Specifically, a Drude weight and coherence temperature scaling as
$\delta^{2}$ are also found. The specific heat coefficient is found to scale
as $1/\delta $ in this case, in contrast to the present work. The reason for
this difference is the existence of a residual  entropy in the undoped
spin-liquid phase of our model. However, the temperature dependence of the
specific heat at the critical point is found to be $\sqrt{T}$ in both cases. 

Finally, we briefly discuss the possible instabilities of the metallic
paramagnetic phase discussed in this paper. It can actually be checked that
for a given $J/t$, a low temperature and low doping regime exists in which an
instability to phase separation is found, signaled by a negative
compressibility. This is quite easily explained on physical basis for a given
realization of the exchange couplings : the holes will tend to cluster in
regions with ferromagnetic bonds in order to maximize kinetic energy. A proper
treatment of this phase separated regime should take into account longer range
Coulomb repulsion. In the infinite connectivity limit, an additional term $V
\sum_{ij} n_{i}n_{j}$ in the Hamiltonian reduces to a Hartree shift $\mu + V
<n>$ of the chemical potential (thus the compressibility reads  :
$\kappa^{-1}_{V}= \kappa^{-1}_{V=0} + V$), so that the phase separation
boundary can be continuously tuned as a function of $V$.                 
In future work, we are planning to consider other possible instabilities of
this model. The issue of spin glass ordering \cite{Rozenberg} 
does not arise for the $M=\infty$ fermionic representation considered
 in this paper \cite{spinfluid}, but spin
glass phases are indeed present for $M=\infty $ for bosonic representations
with high enough 'spin' \cite{OPAGsf}. Even in the fermionic case, first order
corrections in $1/M$ are likely to restore a regime of spin glass ordering. 
Finally, an open issue is that of possible pairing instabilities of the
metallic phase towards a superconducting state.

\subsection{Relevance to cuprate superconductors}

In this section, we would like to present arguments suggesting that the 
problem studied in this paper may be relevant for the understanding of 
some of the striking aspects of the normal state of 
cuprate superconductors. The line of arguments relies on three sets of 
experimental observations:
\begin{itemize}
\item The experiments reported in \cite{ando1,boeb}, in which a $61T$  
magnetic field is used to suppress superconductivity strongly 
suggest that the ground-state of \LSCO\ is actually an {\it insulator}, 
up to Sr-doping of about $x\simeq 0.16$ corresponding to the highest 
$T_c$. This is true even in samples having large values of $k_F l$, 
making weak localization effects an unlikely explanation of the logarithmic 
upturn of {\it both} $\rho_{ab}$ and $\rho_c$ observed at low temperature. 
Insulating behaviour is no longer found in overdoped samples.

\item At very low doping in the \LSCO\  compounds, a low temperature 
spin glass phase is found for $x>0.02$ \cite{sglsco}, in agreement with theoretical 
arguments \cite{aha} suggesting that localized holes induce locally a strong frustration 
in the magnetic exchange. This localization of the carriers induces a strong 
upturn of $\rho_{ab}$ at low temperature in these samples, first in 
a logarithmic manner followed by an activated behaviour. Nevertheless, the 
high-temperature behaviour of the resistivity in these samples is quite 
similar to that found close to optimal doping.
   
\item Inelastic neutron scattering reveals peculiar 
low-energy spin excitations for all underdoped samples,
quite different in nature from spin waves \cite{nlabaco}-\cite{sidis}.  
For very low doping, these excitations occur in a remarkably low 
energy range, on the scale of $10\,meV$, 
distinctly smaller than $J_{AF}$.
In a restricted range of frequency and temperature, the energy scale for 
these excitations is actually set by the temperature itself and $\omega/T$
scaling applies \cite{nlabaco,nlscosg}
These excitations, which are present in a wide range of temperature 
(much above the 
freezing transition mentioned above) and in the whole underdoped 
regime, correspond 
to a {\it slower spin dynamics} than in a Fermi liquid, 
as is also clear from  
the non-Korringa behaviour of the copper NMR relaxation time. 
Similar observations have been made in the \YBCO compounds 
\cite{reviewnybco}. This 
is particularly clear when a small amount of $\mbox{Zn}$ substitution is 
used to suppress superconductivity \cite{sidis} 
(we note that this simultaneously 
opens up again a region of glassy behaviour at low temperature for a rather 
wide range of oxygen content \cite{alloul}).

\end{itemize}

\noindent
In our view, these observations suggest that, in the absence of 
superconductivity, a $T=0$ metal-insulator transition occurs at some 
critical value of the doping $x=x_{MI}$. This transition 
 might be rather close to 
optimal doping in \LSCO \cite{boeb}.
For $x>x_{MI}$, the incipient ground-state 
is a Fermi liquid, corresponding to the overdoped regime. For $x<x_{MI}$, the 
ground state is a Mott-Anderson 
insulator in which holes are localised at $T=0$. 
At very small $x$ ($0.02<x<0.05$),
the mechanism for this hole localization has been studied 
in Ref. \cite{Gooding} and involves both the freezing of hole motion due to
the antiferromagnetic spin background and impurity effects.
This localisation induces strong frustration 
in the local exchange, in agreement with the arguments of Ref.\cite{aha}. 
As a result, this insulator will have a glassy nature at $T=0$ for low 
doping, as indeed found in \LSCO. Beyond $x=0.05$ however, the onset 
of superconductivity has prevented up to now an investigation of the 
low-temperature properties of this incipient insulating ground-state and the
origin of the observed localization is still an open problem. It may be 
that the insulator looses its glassy character at some critical doping 
$x_{g}$ below $x_{MI}$, or that the two critical points actually coincide
($x_{g}= x_{MI}$).
 
\begin{figure}
\[
\fig{12cm}{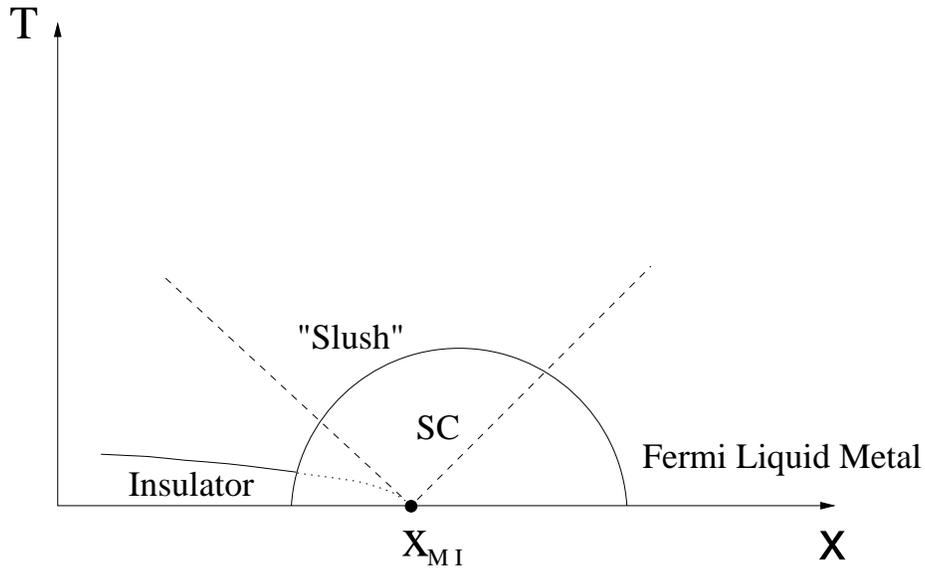} 
\]
\vskip 1cm
\caption{ \label{Phenomenofig}
Schematic crossover diagram for cuprates illustrating : i) the existence of a
metal-insulator transition  as a function of doping at $T=0$ and ii) the
possible relevance of our model to the corresponding quantum critical regime
(other features like the pseudogap and the N{\'e}el temperature have not been
depicted)}
\end{figure}

As temperature is raised, the holes become gradually mobile. This quickly 
destroys the glassy ordering, leaving the system in a ``slushy'' state of 
mobile holes and spins.
Neutron scattering and NMR experiments show that the spin dynamics in this
regime is 
much {\it slower} than in a Fermi liquid state, with local spin correlations 
decaying (in some time range) as $1/t$ 
(corresponding to a high density of low-energy spin excitations 
$\chi''(\omega)\propto\mbox{const.}$ in some frequency range).
We view the model studied in this paper as a simplified description 
of such a slushy state of spins and holes, valid in the high-temperature
quantum critical
regime associated with the transition at $T=0$, $x=x_{g}$ (or $x_{MI}$),
as depicted schematically on Fig.\ref{Phenomenofig}. 
Indeed it is a model of a doped Mott insulator with strong frustration, in
which the effect of quantum disordering the glassy insulating state is 
mimicked by taking the large-$M$ limit. Fluctuations in the transverse
 components of the spin may actually  be an essential 
ingredient in the disordering process, and this is precisely the effect which 
is emphasized in the large-$M$ limit and produces the high density of 
low-energy spin excitations.
 
Of course the present model is highly simplified and is meant to
retain only the interplay of Mott localization with that of frustration in the
magnetic exchange constants. As such, it does not 
include several important physical aspects of the actual materials, most notably:

\begin{itemize}
 
\item i) The fact that frustration is a consequence of hole localisation
at low temperature \cite{aha} (in our model frustration is introduced by hand).
 
\item ii) Localization of carriers by disorder (as a consequence of both
i) and ii),  the metal-insulator transition occurs
at zero-doping in our model).
 
\item iii)  The average antiferromagnetic component $J_{AF}$ of the 
exchange has not been included (in that 
sense we are dealing with a strong frustration limit $J\gg J_{AF}$).
This could be corrected for by reintroducing $J_{AF}$ in a mean-field manner, 
leading to :
$\chi(q,\omega)^{-1} \simeq \chi_{loc}(\omega)^{-1} + J_{AF} \Delta(q)$
where $\Delta(q)$ is the Fourier transform of the nearest neighbour 
connectivity matrix on the lattice. We note that this formula produces a
susceptibility peaked at the antiferromagnetic wave vector, with a correlation
length of the order of the lattice spacing, while all the non trivial dynamics
comes from local effects.

\end{itemize}

\noindent
For these reasons, the present model is unable to address the question of the
precise nature of the incipient insulating ground-state of underdoped
materials (or of the low-temperature pseudogap regime associated with it),
even though the remarks made above point towards a phase separated regime. 
Various proposals have been made in the literature regarding this issue. One
of the most widely discussed is the ``stripes'' picture, in which there is
phase separation between the doped holes and the spins into domain-wall
like structures. We note that as long as the holes remain confined in these
structures, the mechanism of Ref. \cite{aha}
implies the existence of ferromagnetic bonds in the
hole-rich region, as indeed found in numerical calculations
\cite{Gooding,white,fleck}.
As temperature or doping is raised, a melting transition of the stripe
structure takes place, and the model of a ``spin-hole slush'' introduced
here may become relevant in the associated quantum critical regime.  

Keeping these caveats in mind, we comment on the  
comparison between our findings and some aspects of the normal state of 
cuprates in the regime depicted schematically on Fig.\ref{Phenomenofig}:

\begin{itemize}

\item {\it Low-energy coherence scale}

The present model yields a remarkable
suppression of the low-energy coherence scale of
a doped Mott insulator in the presence of frustrating exchange couplings.
We find this scale to be of order 
$\epstar = (\delta t)^2/J = J (\delta/\delta^*)^2$
instead of the naive (Brinkman-Rice) estimate $\delta t$. We note that, with
$t/J\simeq 5$, and $J \simeq 1200 K$, this scale
 can be as low as a few hundred degrees.
If relevant for cuprates, this observation suggests that
 the normal state properties
may well be associated, over an extended (high-) temperature regime,
with {\it incoherent behaviour} characteristic of a quantum
critical regime dominated by thermal effects. 
We note however that the present model, as any model in which low-energy 
excitations are local in character, would lead to a large effective mass at
low temperature, directly proportional to $1/\epstar $. In cuprates,
additional physics  sets in at lower temperature (cf  { iii) } above)
which quench the corresponding  entropy, leading to the experimentally 
observed moderate effective mass \cite{cooper}.

\item {\it Photoemission} 

In the incoherent ``slushy'' regime 
$T>\epstar$, we find a single
particle Green's function decaying as $1/\sqrt{\omega}$
(and an associated single-particle lifetime $\Im\Sigma\propto\sqrt{\omega}$),
leading to a markedly
non-Fermi liquid tail of the photoemission intensity. It is worth noting that
precisely this form has been recently shown to provide a rather good fit to the
high-frequency part of the photoemission lineshape above the pseudogap
temperature in underdoped $Bi_2Sr_2CaCu_2O_{8+x}$ \cite{chubu}. 
It has been recently argued that the $1/\sqrt{\omega }$ behaviour also holds
in the strong coupling limit of antiferromagnetic spin fluctuation theories
\cite{overdamped}.
 
\item {\it Resistivity and Optical conductivity}

We would like to emphasize again the mechanism which yields
a linear resistivity in the incoherent regime of our model, 
starting from a single-particle self energy behaving
as $\sqrt{\omega}$. This holds when scattering is {\it local} and
{\it incoherent} so that the
effective quasiparticle bandwith (dispersion) can be neglected in comparison
to lifetime effects. In this limit, conductivity should be thought of in real
space  as a tunneling process between neighbouring lattice sites.
This mechanism has a higher degree of generality than the
specific model considered in this paper, and should also apply to other models
in which the same $\sqrt{\omega }$ power-law behaviour of the self-energy 
holds, such as the model of Ref.\cite{overdamped}.
This model has been proposed in connection with the normal state properties
of underdoped cuprates above the pseudo-gap temperature \cite{chubu}.  

We note that the magnitude of the linear resistivity in this incoherent
regime is larger or comparable to the Mott limit ($a h/e^2$), 
as is actually the case over
a rather extended high temperature regime in underdoped \LSCO \cite{takagi}
and is a quite general feature of `bad metals'.

Regarding optical conductivity, the form we have obtained is quite 
similar to the Marginal Fermi Liquid one, which has been shown 
\cite{elihu} to provide a very good fit to the data of e.g.
Ref.\cite{nb1,nb2}.  

\item {\it Neutron scattering}

Neutron scattering experiments on non-superconducting 
\LBCO\  \cite{nlabaco} and \LSCO\  with $x=0.04$ \cite{nlscosg,reviewnlsco} 
have revealed spin excitations which are centered at the wavevector 
$Q=(\pi,\pi)$ with a rather large momentum width. The frequency 
dependence of these excitations display $\omega/T$ scaling and have 
been successfully fitted by scaling forms very similar to the one found 
in the present model \cite{nlabaco,nlscosg}. 
At higher $Sr$ concentration, one of the most 
notable feature of the neutron scattering results is the appearance of 
sharp peaks at incommensurate wavevectors. It is likely however that these 
peaks only carry a small fraction of the total spin fluctuation intensity, 
as suggested in particular by comparison to NMR data. A broad, weakly 
$q$-dependent contribution most probably persists up to high temperature, 
carrying a large part of the total weight, and hard to distinguish from 
''background'' noise in neutron experiments \cite{personal}. 
In \YBCO, suppression of superconductivity by $Zn$ doping allow to 
investigate the spin dynamics of the normal state down to low 
temperature \cite{sidis,reviewnybco}. Apart from a very low temperature 
quasi-elastic peak (associated with spin freezing into  
spin-glass like  order), neutrons scattering results for $y=0.39$ 
reveal a strong enhancement of low-frequency spin fluctuations 
at low temperature, with a distinctively low energy scale and 
a strong temperature dependence down to very low temperature 
(compatible with $\omega/T$ scaling in a limited range). These features 
are qualitatively similar to the low-energy excitations found in the 
present model. There is furthermore experimental evidence 
\cite{sidis} that these 
low-energy excitations are associated with the disordering of the 
spins by transverse fluctuations, as in our model.  

\end{itemize}
\acknowledgements
We are most grateful to Subir Sachdev for useful correspondence and remarks.
We acknowledge useful discussions with  our experimentalist colleagues:
N. Bontemps, G. Boebinger, H. Alloul and particularly with
P. Bourges, G. Colin, Y. Sidis, S. Petit, L.P. Regnault 
and G. Aeppli on their neutron scattering results.
We also thank  A. Millis, C. Varma and particularly E. Abrahams
for their suggestions and comments. 
Part of this work was completed during stays of A.G
at the I.T.P, Santa Barbara (partially supported by
NSF grant PHY94-07194) and at the Rutgers University Physics Department.

\appendix

\section{Derivation of the saddle-point equations}
\label{apxspeq}

In this appendix, some further details on the derivation of the 
saddle-point equations in the large-$M$ limit for the 
single site model defined by Eqs.(\ref{action2},\ref{defDetR})
are  provided.
In the following, we will drop the index $\alpha$ in 
$G^{ab}(\tau -\tau')\equiv
-\frac{2}{M} <(f^{a}_\alpha b^{\dagger a})(\tau)
(f^{\dagger b}_\alpha b^{b})(\tau')>$. 
In  (\ref{defDetR}) the brackets denote the average
with the action specified in subscript.
As the action $S$ is invariant under translations in imaginary time
and under the action of $SU(M)$ (the rotation invariance for $M=2$),
$C $ and $R$ take the following form :
\begin{eqnarray}\label{SuNInvariance}
\nonumber
C^{aa}_{\alpha\alpha}(\tau,\tau')&=&
\frac{M}{2} G^{aa}(\tau -\tau')\\
 R^{ab}_{\alpha \beta \gamma \delta }(\tau,\tau') &=&
-  \delta_{\alpha \delta } \delta_{\beta \gamma } R^{ab}(\tau -\tau')
+ \delta_{\alpha \beta }\delta_{\gamma \delta }
\widetilde{R}^{ab}(\tau -\tau')
\end{eqnarray}
The quartic term in $f$ in (\ref{action2}) is  decoupled using a bi-local
 field $P^{ab}(\tau,\tau')$. Using the expression of the spin operator
Eq. (\ref{ExprSf})  and the change of variable :
\begin{equation}\label{CondensationBoson}
 b(\tau )=\sqrt{\frac{M}{2}} \phi(\tau)
\end{equation}
the single-site partition function can be rewritten as : 
\begin{equation}\label{DecouplageGaussien}
\Zeff = \int \gd \phi^{\dagger}\gd \phi \gd\lambda \gd P
  e^{-M S_{1}-\widetilde{S_{1}}}
\end{equation}
with the actions :
\begin{eqnarray}\label{action3}
\nonumber
S_{1}&=&
 \frac{1}{2} \int \! d\tau \sum _{a}
\phi ^{\dagger a}(\tau)\partial_{\tau }\phi^{a}(\tau)
-  \log Z_{0} + \frac{J^{2}}{2}  \sum_{ab} \int\!\!\!\int d\tau d \tau'\,
  R^{ab}(\tau - \tau') P^{ab}(\tau,\tau') P^{ba}(\tau',\tau)\\
\widetilde{S_{1}} &=& \frac{J^{2}}{2} \sum_{a} \left( \int \! d\tau
\left( 1-\phi^{\dagger}(\tau) \phi(\tau)  \right)
\right)^{2}
\end{eqnarray}
In this expression, $Z_{0}$ is defined by 
\begin{mathletters}
\begin{equation}\label{DefZ0}
Z_{0}[\phi,P,\lambda ]\equiv  \int \gd f^{\dagger } \gd f
 e^{- S_{00}[\phi,P,\lambda,f]}
\end{equation}
with :
\begin{eqnarray}\label{action4}
\nonumber
&&S_{00}[\phi,P,\lambda,f] = \sum_{a} \int d\tau
 f^{\dagger a}(\tau)(\partial_\tau -\mu) f^a(\tau) \\
\nonumber
&& + i \int\!  d\tau  \sum_{a}\lambda^{a} (\tau)
\left(
 f^{\dagger a}(\tau) f^a(\tau)
+ \frac{\phi^{\dagger a}(\tau)\phi^a(\tau) - 1}{2}
\right) \\
\nonumber &&
- J^2 \sum_{a,b } \int\!\!\!\int  d\tau d \tau'\,
 R^{ab}(\tau-\tau') P^{ab}(\tau,\tau') f^{\dagger b}(\tau') f^{a}(\tau) \\
&& + t^{2}\sum_{a}\int\!\!\!\int d\tau d \tau'\,
 f^{\dagger a}(\tau) \phi^a(\tau) G^{a}(\tau-\tau')
 \phi^{\dagger a}(\tau')  f^a(\tau')
\end{eqnarray}
\end{mathletters}
In the limit $M\rightarrow\infty$, $\Zeff$ is controlled by a saddle point
 with respect to $P^{ba}(\tau',\tau)$, $\lambda (\tau )$ and $\phi(\tau )$.  
We assume a {\it condensation of the boson}~:
after the change of variable (\ref{CondensationBoson})
$\phi $ is taken to be a finite constant at the saddle-point :
$ \phi_{sp}(\tau) =\sqrt\delta$ and $\lambda $ is static : 
$i\lambda_{sp}(\tau )=\lambda_{0}$. 
Moreover, in this limit the correlation functions of $f$  are given by the
average with the action $S_{00}$ taken for these values of $P,\lambda,\phi$. 
As $S_{00}$ is quadratic in $f$ (and the boson is condensed),
 the model is completely solved in this limit 
as soon as $G_{f}$ has been calculated.
The saddle point equations are given by the 
minimisation of $S_{1}$  with respect to $P^{ba}(\tau',\tau)$,
$\lambda (\tau )$ and $\phi(\tau )$ respectively, which leads to :
\begin{eqnarray}\label{sp1}
\nonumber
P^{ab}(\tau,\tau') &=& - <f^b(\tau')f^{\dagger a}(\tau)>_{S_{00}} \\
\nonumber
1 &=& \delta - 2 <f^{a}(\tau) f^{\dagger a}(\tau)>_{S_{00}}  \\
\lambda_0 \sqrt\delta  &=& -2 t^2 \delta^{\frac{3}{2}}
\int_{0}^{\beta} d \tau G^{aa}(\tau)G^{aa}(-\tau)
\end{eqnarray}
and finally gives Eq.(\ref{EqBase1},\ref{EqBase2},\ref{EqBase3},\ref{EqBase4})
 given in the text.

\section{Luttinger Theorem}\label{Lutt}
In order to find the volume of the Fermi surface in the interacting system, we
proceed along the lines of Ref.\cite{AGD} and we observe that the auxiliary 
fermion self-energy can be obtained as the functional derivative of 
the following functional:
\begin{equation}
\Phi = J^{2}\int d t \,
\left( G_{f}^{F}(t) G_{f}^{F}(-t) \right)^{2}\,\,\,,\,\,\,
\Sigma_{f}^{F}(t)= \frac{\delta \Phi }{\delta G_{f}^{F}(-t)}
\end{equation}
The number of particles reads:
\begin{equation}
{{1-\delta}\over{2}}=
\int_{-\infty }^{\infty } 
\frac{d\omega }{2 i \pi }
  G_{f}^{F}(\omega) \,\,e^{i \omega  0^{+}}
\end{equation}
and we use the identity:
\begin{equation}\label{lutt3}
G_{f}^{F}(\omega) = 
\frac{\partial }{\partial \omega }\int_{-\infty }^{\infty } d\epsilon \,
D(\epsilon)
\ln \left( \omega +\mu -\lambda_{0} - \Sigma_{f}^{F}(\omega) -\delta\epsilon 
\right)
+ \frac{\partial \Sigma_{f}^{F}(\omega )}{\partial \omega } G_{f}^{F}(\omega )
\end{equation}
Using the invariance of the Luttinger-Ward functional under a shift of all
 frequencies 
($G(\omega)\rightarrow G(\omega+\Omega)$), the integral of the last term 
vanishes:
\begin{equation}\label{lutt4}
\int_{-\infty }^{\infty } d\omega
\frac{\partial \Sigma_{f}^{F}(\omega )}{\partial \omega }G_{f}^{F}(\omega )=0
\end{equation}
and the integral of the first term can be explicitly calculated by
transforming to retarded Green's functions (denoted by $G_{f}^{R}$) in the 
following manner:
\begin{equation}
\frac{1-\delta}{2}= 
\int_{-\infty }^{\infty} d\epsilon \,D(\epsilon ) 
\left[ -
\int_{-\infty }^{\infty }  \frac{d\omega }{2 i \pi }
\partial_{\omega } \ln  G_{f}^{R}(\epsilon,\omega) e^{i\omega 0^{+}}
+\int_{-\infty }^{0}  \frac{d\omega }{2 i \pi }
 \partial_{\omega } \ln  \left(\frac{G_{f}^{R}(\epsilon ,\omega )}
{\overline{G_{f}^{R}(\epsilon ,\omega )}}\right) 
\right]
\end{equation}
As $G_{f}^{R}$ has no pole nor zeros in the upper half plane, 
the first integral can be closed there and vanishes.
Hence we have : 
\begin{equation}\label{lutt2}
\frac{1-\delta }{2}= \int_{-\infty }^{\infty } d\epsilon \,D(\epsilon )
\Theta \left(\mu -\lambda_{0} - \Sigma_{f}(i 0^{+}) -\delta\epsilon \right)
\end{equation}
From Eq.(\ref{lutt2}) and the definition of 
$\mu_0$, we finally obtain:
\begin{equation}
\mu(T=0)-\lambda_0(T=0)-
\Sigma_f(\omega=0,T=0) = \delta\mu_0(\delta)
\end{equation}
which is the desired relation and insures that Luttinger theorem holds 
in the presence of both the constraint and the magnetic scattering. We also 
checked that this property is verified in our numerical calculations at $T=0$.

\section{Numerical method}\label{AppendiceNum}

In this appendix, we explain the main steps that we followed in
solving numerically the saddle point equations Eqs.(\ref{EqBase}).

\subsection{Computation of the Green function $G_{f}(\omega,T)$}

The calculation of the Green's function is divided in two steps. First
a Matsubara frequency/imaginary time algorithm is used in an iterative
manner in order to find the value of the chemical potential $\mu$ and 
Lagrange multiplier $\lambda$ for a given doping $\delta$, interaction
strength $J/t$ and temperature. Convolutions are calculated using a 
FFT algorithm and a simple iteration is used: starting from a given
$G$, a self-energy is obtained which is then reinjected into the
expression for $G$ until a converged set $(G,\Sigma)$ is reached (for
given values of $\mu,\lambda$).   
A second routine uses the previous one to adjust $\mu $ and $\lambda_{0}$ 
in order (\ref{EqBase3},\ref{EqBase4}) to be satisfied.

Once the imaginary-time Green's function and values of $\mu,\lambda$
have been found using this imaginary time algorithm, a different
algorithm is used to obtain {\it real frequency} Green's functions and
spectral densities. This is done in the following manner. We 
consider a {\it finite-temperature} generalisation of the 
Green's function with the Feynman prescription: 
\begin{equation}\label{DefFeynGreenT}
G(\omega ) \equiv \left(1-n_{F}(\omega )\right) G^{R}(\omega)  +
n_{F}(\omega ) \overline{G^{R}(\omega )}
\end{equation}
which reduces to the usual Green's function $G_{F}$ at $T=0$.
We now define: 
\begin{equation}\label{Defsigtilde}
\widetilde{\Sigma }(t) \equiv  J^{2} G(t)^{2}G(-t)
\end{equation}
where $t$ is the real time.
Expressing both $\widetilde{\Sigma}$ and $\Sigma^{R}$ as integrals of the 
spectral density with the spectral representation of $G^{R}$, we obtain, 
after some calculations (R is a superscript denoting {\it retarded}
quantities) :  
\begin{equation}\label{RelFondNum}
\widetilde{\Sigma }(\omega ) = \Sigma^{R}(\omega ) + 
2i \pi  J^{2} \int \!\!\! \int d \omega_{1} d \omega_{2}
n_{F}(\omega_{1}) \rho(\omega_{1}) n_{F}(\omega_{2}) \rho(\omega_{2})
n_{F}(\omega -\omega_{1} - \omega_{2}) \rho(\omega -\omega_{1} - \omega_{2})
\end{equation}
where $\rho =- \frac{1}{\pi} \Im G^{R}$ and $n_{F}$ is the Fermi factor.

At $T=0$, Eq. (\ref{RelFondNum}) shows that 
$\widetilde{\Sigma }$ simply coincides with $\Sigma^{F}$, 
the usual self-energy at $T=0$, with 
the Feynman prescription and that Eq. (\ref{EqBase}) 
can be rewritten as, at $T=0$ : 
\begin{eqnarray}\label{EqBaseTnul}
\nonumber
(G^{F}_{f}(\omega ))^{-1}
&=& \omega +\mu-\lambda_0-(t\delta)^2 G^{F}_{f}(\omega) 
-\Sigma^{F}_{f}(\omega )\\
\Sigma^{F}_{f}(t) &=&  J^2 (G^{F}_{f}(t))^{2} G^{F}_{f}(-t)
\end{eqnarray}
together with the equation corresponding to (\ref{EqBase3},\ref{EqBase4}).
Note the change of the sign in front of $J^{2}$.
This form was used in (\ref{Equationgf}). From (\ref{EqBaseTnul})
one can write an algorithm for the computation of $G^{F}$ in the $T=0$
 formalism, similar to the computation in imaginary time.
Of course, as our large-$M$ limit performs a resummation of the
perturbation theory, one can also 
obtain these equations with the diagrammatic rules, but these rules
do not apply at finite temperature to the Green's function $G$ in a
systematic manner.

A finite temperature, we use the following iterative algorithm :
Starting from $G^{R}$, we get $G$. We then obtain $\widetilde{\Sigma }$ 
by direct convolution in  real time and the second term of the r.h.s  of 
(\ref{RelFondNum}) by a double convolution of $n_{F}\rho$. Hence we obtain
$\Sigma^{R}$ and go back to $G^{R}$ with Eq. (\ref{EqBase}).

As a starting point of the iteration, in order to speed up
convergence,  
we take an analytic continuation of the solution
in imaginary time, obtained by a standard Pade approximation. 
Note that the parameter $\mu$ and $\lambda_{0}$ are fixed 
in this iteration
on the real axis, since they have been 
calculated before in the Matsubara formalism. 

As soon as the Green function has been obtained, some 
other quantities are
straightforwardly calculated from the formula given in the text.
In particular, $\chiloc''$ is expressed as a convolution.

However the computation of the uniform susceptibility $\chi $ (considered in
Appendix \ref{Susceptibilite}) is more
involved : we solve  Eq.(\ref{SuscepUnif}) for $g$ by another iterative loop
analogous to the previous ones.

The scaling function describing the effect of the doping at $T=0$, is 
computed from (\ref{Equationgf}) by an iterative algorithm similar
to the previous ones.

\subsection{Computation of the resistivity}

We also give some useful details on the numerical calculation of 
the frequency-dependent resistivity. There, it is very convenient to 
integrate analytically over $\epsilon $
in (\ref{ConducBase}) using (\ref{DefRhoc}) and the relation :  
\begin{equation}
\int d \epsilon \,\, \frac{D(\epsilon )}{A(\nu) -\epsilon } = G_{f}(\nu)
\end{equation}
We thus obtain (the Green function is the retarded one) :
\begin{equation}\label{ConducReduit}
\Re\sigma (\omega ) = \frac{\delta ^{2}}{8 \pi ^{2}} \int d\nu  \Re 
\left[ \frac{G_{f}(\nu +\omega) - G_{f}(\nu)}{A(\omega +\nu) - A(\nu)}
- \frac{G_{f}(\nu) - \overline{G_{f}(\nu+\omega)}}
{A(\nu) -  \overline{A(\nu+\omega )}}
 \right] \frac{n_{F}(\nu ) - n_{F}(\nu +\omega)}{\omega}
\end{equation}
For the dc-conductivity, Eq.(\ref{ConducReduit}) simplifies to :
\begin{equation}\label{ConducStatReduit}
\sigma (T) = \frac{ 1 }{32 \pi ^{2} T} \int d \omega   \,\,
\left[\Re \left(\frac{1}{\delta^{2} G_{f}^{2}(\omega) -1 } \right) 
 - \frac{1}{\delta^{2} |G_{f}(\omega)|^{2} -1 }\right]
\frac{1}{\cosh ^{2} \frac{\beta \omega  }{2}}
\end{equation}
In both case the integrals are computed simply by transforming them into
a Riemann sum.

\section{Scaling analysis}\label{AppFonctionsEchelle}

In this Appendix, some details  about the thermal scaling  analysis of
Section \ref{SectCrossOverTherm} in the spinfluid regime are provided.
As explained in the text, in this regime $\epstar$ disappears from the
thermal scaling functions and thus the calculation can be performed in 
the undoped model $\delta =0$. In this Appendix, $G_{f}$ and $\Sigma_{f}$ will
denote the thermal scaling function of these quantities.
They satisfy the scaled saddle point equation 
$G_{f}(\omega/T)^{-1} = - \Sigma_{f}(\omega/T)$.
The calculation is very similar to the low temperature, low frequency analysis
made in \cite{OPAGKondo} so we here just give the main steps of the analysis.  
We first check that the scaling behaviour (\ref{GEchelle}) in imaginary time 
solves the saddle point equation (for $\delta =0$) using the 
Fourier formulas (which follows from \cite{Grad} 3.631) : 
\begin{mathletters}
\label{fourier}
\begin{equation}
G_f(i \omega_n) =  - { i
\pi^{\frac{1}{4}} (JT)^{-\frac{1}{2}} (-1)^n \Gamma(\frac{1}{2}) 
\over 
\Gamma\left(\frac{3}{4} - {\omega_n\beta \over 2 \pi} \right)
\Gamma\left(\frac{3}{4} + {\omega_n\beta \over 2 \pi} \right)
} 
\end{equation}
\begin{equation}
\Sigma_f(i \omega_n) = - 
{i  \pi^{\frac{3}{4}} \sqrt{JT} (-1)^n 
\Gamma(-\frac{1}{2})
\over
\Gamma\left({1\over 4} - {\omega_n\beta \over 2 \pi} \right)
\Gamma\left({1\over 4} + {\omega_n\beta \over 2 \pi} \right)
 } 
\end{equation}
\end{mathletters}
where $\omega_{n} = (2n+1)\pi T$ are the Matsubara frequencies.
One can then show that (\ref{ScalingFormRho}) and (\ref{ScalingFormSigma})
are the scaling function on the real axis using the following method.
First we use  
\begin{equation}
G_{f}(\tau) = - \int_{-\infty}^{+\infty} {e^{-\tau \varepsilon}\over 
1 + e^{-\beta\varepsilon}} \rho_{f}(\varepsilon) \,\, d \varepsilon 
  \,\,\,\,\,  0\leq\tau\leq\beta
\end{equation}
and  
\begin{equation}
\label{fourier2}
\int_{-\infty}^{+\infty} d t
 \left( {\pi \over \cosh (\pi t)}\right)^\Delta e^{-itu}
 = (2\pi)^{\Delta-1} 
{
\Gamma\left({\Delta\over 2} + {iu\over 2 \pi}\right)
\Gamma\left({\Delta\over 2} - {iu\over 2 \pi}\right)
\over 
\Gamma(\Delta)}
\,\,\,\,\,\,\, 
 \left\{{ 0<\Delta<1 \atop u \hbox{ real}}\right.
\end{equation}
(See formula 3.313.2 of \cite{Grad}).
We then deduce the full Green function (and thus $\Sigma_{f}$) by performing 
 the Hilbert transform 
of $\rho_{f}$ using (\ref{fourier2}) again and 
\begin{equation}
\label{fourier3}
\int_{0}^{+\infty} d x
{e^{iz x}\over 
 \left( \sinh {\pi x\over \beta}\right)^\Delta }
 = 2^{\Delta-1}{\beta\over \pi} 
{
\Gamma\left({\Delta\over 2} - {i\beta z\over 2 \pi}\right)
\Gamma\left( 1- \Delta \right)
\over
\Gamma\left(1-{\Delta\over 2}-i{\beta z\over 2\pi}\right)
}
\,\,\,\,\,\,\,\,\,\,
\left\{{ 0<\Delta<1 \atop z \hbox{ real}}\right.
\end{equation}
(See formula 3.112.1 of \cite{Grad}).

\section{Uniform susceptibility}\label{Susceptibilite}

In this appendix, we briefly explain how to calculate the {\it uniform
susceptibility $\chi$} and analyse its behaviour at small temperature in
the undoped model $\delta =0$.

\subsection{Effect of a  magnetic field}\label{EqGeneChampMag}
 The magnetic field is  introduced in the
$SU(M)$ Hamiltonian (\ref{Ham1}) in the following way :
\begin{equation}\label{DefChampMag}
\delta H = - h \left(f^{\dagger}_{1}f_{1} - f^{\dagger}_{2}f_{2}\right)
\end{equation}
This formula clearly reduces to the usual one for $M=2$. Here, only color
 1 and 2 are coupled to this field but note that this choice is not unique
though convenient for our calculation (more generally the magnetic field must
be coupled to an element of a Cartan subalgebra of SU(M)).
 The large $d$ and large $M$ limit computation is similar
to the zero field one explained previously, although it is more
involved. 
A simplification occurs in this double limit : due to the fact that only
2 colors over $M$ are coupled to $h$, the Green function $G_{f}^{i}$ for 
colors $ i > 2$ are solutions of the zero-field equations (\ref{EqBase}).
Moreover we find $G^{1}_{f}=G^{h}$ and $G^{2}_{f}=G^{-h}$ where
 $G^{h}$ is given by :
\begin{mathletters}
\begin{equation}\label{EqBase5}
\left( G^{h}_{f}(i\omn) \right)^{-1} = i\omn+\mu -\lambda_0 +h 
-(t\delta)^2 G_{f}^{h}(i\omn) 
-\Sigma_{f}^{h}(i\omn)
\end{equation}
\begin{equation}\label{EqBase6}
\Sigma_{f}^{h}(\tau) \equiv  -J^2 G_{f}^{h}(\tau ) 
 G_{f}^{h=0}(\tau) G_{f}^{h=0}(-\tau)
\end{equation}
\end{mathletters}
($\mu$, $\lambda_{0}$ and $G_{f}^{h=0}$ are always 
determined by Eqs.(\ref{EqBase})).
The magnetisation  is given by   $m=G_{f}^{1}(0^{-}) -G_{f}^{2}(0^{-})$. 
Let us define $g$ by $G_{f}^{1}- G_{f}^{2} = h g + O(h^{2})$.
 From Eq.(\ref{EqBase5},\ref{EqBase6}), $g$ satisfies :
\begin{eqnarray}\label{SuscepUnif}
 g(i\omn) &=& \frac{K(i\omn)}{(t\delta )^2 - G_{f}^{-2}(i\omn)} \\
K(\tau) &=& 2 \delta (\tau )  + J^2 g(\tau) G_{f}(\tau) G_{f}(-\tau)
\end{eqnarray}
With these notations, the uniform susceptibility is given by
 $\chi =g(\tau =0^{-})$.

\subsection{Low temperature behaviour of $\chi$ for the undoped model}

\begin{figure}
\[
\fig{10cm}{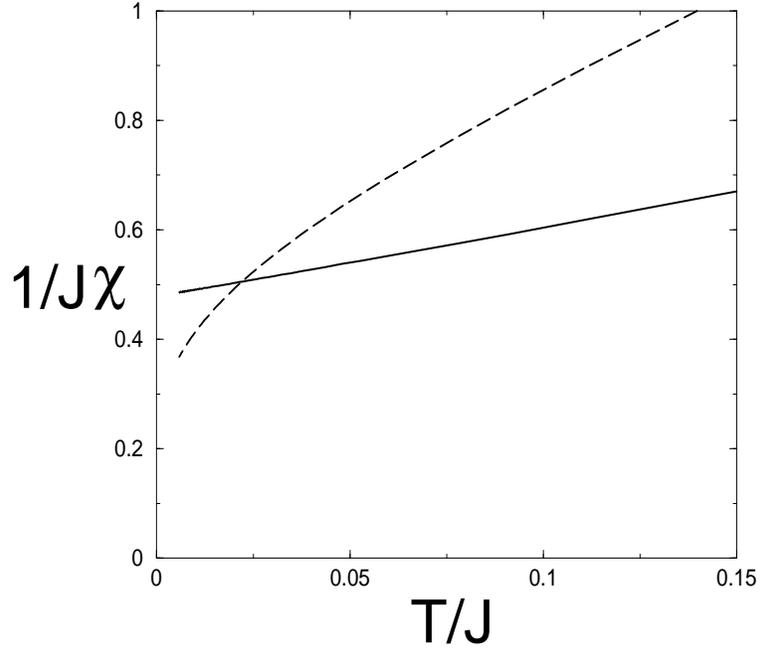} 
\]
\vskip 1cm
\caption{ \label{ChiUnif}
$1/\chi(T)$ (solid line) and $1/\chiloc(T)\sim 1/\ln T$ (dashed line)
 versus $T$ as calculated numerically in the undoped model.}
\end{figure}

In the undoped case, Eq.(\ref{SuscepUnif}) reduces to   
$ g(\omega)= -G_{f}^2(\omega) K(\omega) $ and 
the susceptibility at $T=0$ is formally given by :
\begin{equation}\label{chiTnulle}
\chi (T=0) = \int_{-\infty}^0 g(\omega) d \omega
\end{equation} 
From the low frequency behaviour Eq.(\ref{LowOmBehavG}) we have 
$g(\omega) \sim -i\,\, \const  K(\omega) / \omega $. Thus 
we have to investigate the low-frequency  behaviour of $K$. 
Generalising the Luttinger theorem (as expressed by Eq.(\ref{luttFinal})) 
for the colors 1 and 2, we obtain  (at $T=0$) : 
\begin{equation}\label{luttChamp}
\mu  \pm  h - \lambda_{0} - \Sigma_{f}^{1/2}(\omega =0) =
 \delta \mu_{0}^{\pm h}(\delta) 
\end{equation}
In this equation, $\mu_{0}^{\pm h}(\delta) $ is given by :
\begin{equation}
\int_{-\infty }^{\mu_{0}^{h}(\delta)} d\epsilon \,D(\epsilon ) =
n_{1}^{h}
\end{equation}
where $n_{1}^{h}$ is the number of particles of color 1.
Hence we obtain :
\begin{equation}
2 - \left(\Sigma_{f}^{1}(0) - \Sigma_{f}^{2}(0) \right) =
 \delta \frac{\mu_{0}^{ h}(\delta) - \mu_{0}^{- h}(\delta)}{h}
\end{equation}
Taking first the limit $\delta \rightarrow 0$ and then $h \rightarrow 0 $ 
we obtain finally (as $\mu_{0}$ is bounded by definition):
\begin{equation}\label{cancellation}
K(\omega =0)=0
\end{equation}
Thus the leading low-frequency singularity in $g$ cancels so from 
Eq.(\ref{chiTnulle}) we see that $\chi $ is smaller than $\ln T$ at small
temperature. Strictly speaking we can  not prove from the previous argument 
 that $\chi $
reaches a finite value at zero temperature, but it is a very natural guess
which is moreover very well supported by our numerical calculation as
displayed  in Fig.\ref{ChiUnif}.

\begin{figure}
\[
\fig{10cm}{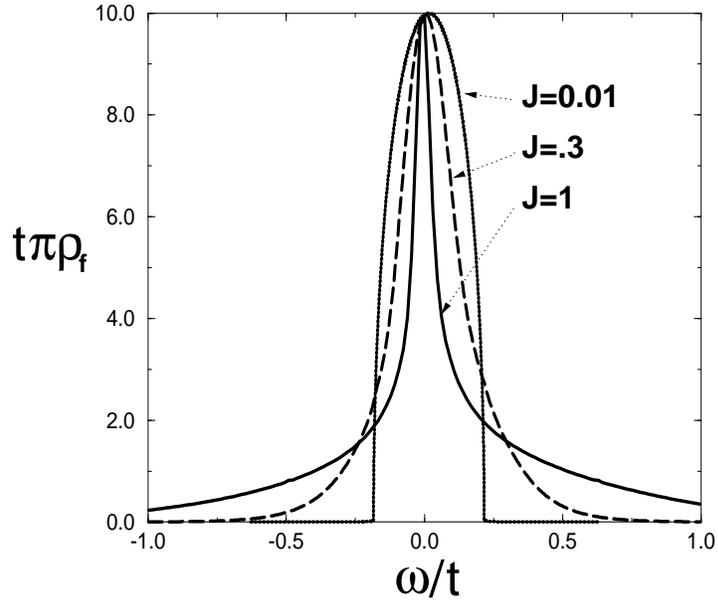} 
\]
\vskip 1cm
\caption{ \label{SpectralFunctions}
The spectral function of the auxiliary fermion as a function of frequency
for a doping  $\delta=0.1$ and three values of $J=0.01,0.3,1$}
\end{figure}

\begin{figure}
\[
\fig{10cm}{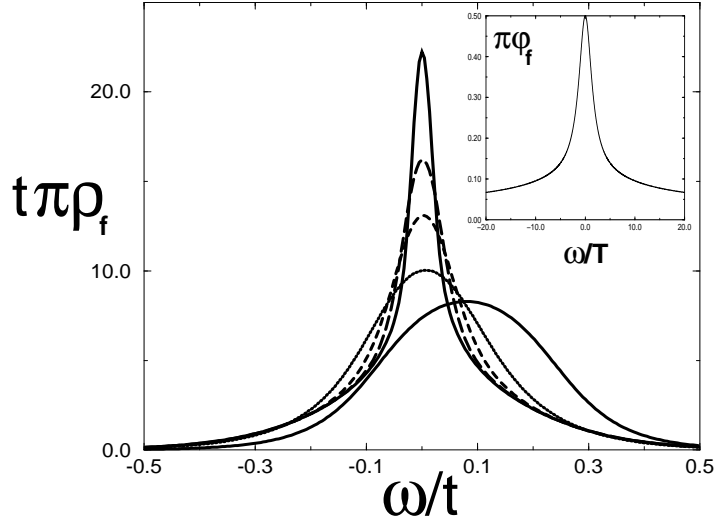} 
\]
\vskip 1cm
\caption{\label{DosFntTemp}
Spectral functions $\pi \rho_{f}(\omega )$ for $\delta =0.04$ and $J/t=0.3$
 (corresponding to $\epstar /J \simeq 1.8\, 10^{-2}$). The different curves
correspond from top to bottom to $T/t = 1/200,1/50,1/25,1/10,1 $. Inset : thermal
scaling function Eq. (\ref{ScalingFormRho}) }
\end{figure}

\begin{references}
\bibitem{warningbosons} This is true
for fermionic representations and for bosonic representations with
small enough ``size'' of the spin.
\bibitem{footsubir} We note that our conventions for Green's functions
are different than those in Ref.\cite{spinfluid}.
\bibitem{footentropy} Only part of the degeneracy is actually removed:
the spin-liquid ground-state turns out to have an extensive entropy in
the large-$M$ limit, as studied in detail in \cite{OPAGsf}.
 b
\bibitem{footnoteFourier}  Note that strictly speaking there is no Fourier
transform on the Bethe lattice. $\k$-dependence should really be interpreted
as a dependence on the single-particle tight binding energy $\epsilon_{\k}$.
\end{references}
\end{document}